\documentclass[iop]{emulateapj}
\usepackage{graphicx,longtable,natbib,hyperref}

\def\arcsec{{\mbox{$^{\prime \prime}$}}}
\def\arcmin{{\mbox{$^{\prime}$}}}
\def\degree{{\mbox{$^{\circ}$}}}

\def\Sw{{\sl Swift}}

\begin{document}
\title{Host Galaxy Properties of the $\Sw$ BAT Ultra Hard X-ray Selected AGN}
\author{Michael Koss\altaffilmark{1,2,3}, Richard Mushotzky\altaffilmark{1}, Sylvain Veilleux\altaffilmark{1}, Lisa M.~Winter\altaffilmark{4,5},Wayne Baumgartner\altaffilmark{2}, Jack Tueller\altaffilmark{2}, Neil Gehrels\altaffilmark{2}, and Lynne Valencic\altaffilmark{2}}
\email{mkoss@astro.umd.edu}
\altaffiltext{1}{Astronomy Department, University of Maryland, College Park, MD, USA}
\altaffiltext{2}{Astrophysics Science Division, NASA Goddard Space Flight Center, Greenbelt, MD, USA}
\altaffiltext{3}{Institute for Astronomy, University of Hawaii, Honolulu, HI, USA}
\altaffiltext{4}{Center for Astrophysics and Space Astronomy, University of Colorado, Boulder, CO, USA}
\altaffiltext{5}{Hubble Fellow}

\begin{abstract}
We have assembled the largest sample of ultra hard X-ray selected (14-195~keV) AGN with host galaxy optical data to date, with 185 nearby (z$<$0.05), moderate luminosity AGN from the ${\sl Swift}$ BAT sample.  The BAT AGN host galaxies have intermediate optical colors ($u-r$ and $g-r$) that are bluer than a comparison sample of inactive galaxies and optically selected AGN from the Sloan Digital Sky Survey (SDSS) which are chosen to have the same stellar mass.  Based on morphological classifications from the RC3 and the Galaxy Zoo, the bluer colors of BAT AGN are mainly due to a higher fraction of mergers and massive spirals than in the comparison samples.  BAT AGN in massive galaxies (log M$_*$$>$10.5) have a 5 to 10 times higher rate of spiral morphologies than in SDSS AGN or inactive galaxies.  We also see enhanced far-IR emission in BAT AGN suggestive of higher levels of star formation compared to the comparison samples. BAT AGN are preferentially found in the most massive host galaxies with high concentration indexes indicative of large bulge-to-disk ratios and large supermassive black holes.  The narrow-line (NL) BAT AGN have similar intrinsic luminosities as the SDSS NL Seyferts based on measurements of  [O \textsc{III}] $\lambda5007$.   There is also a correlation between the stellar mass and X-ray emission. The BAT AGN in mergers have bluer colors and greater ultra hard X-ray emission compared to the BAT sample as whole. In agreement with the Unified Model of AGN, and the relatively unbiased nature of the BAT sources, the host galaxy colors and morphologies are independent of measures of obscuration such as X-ray column density or Seyfert type.  The high fraction of massive spiral galaxies and galaxy mergers in BAT AGN suggest that host galaxy morphology is related to the activation and fueling of local AGN.

\end{abstract}
\keywords{galaxies: active --- X-rays: interactions --- X-rays: galaxies}

\section{Introduction}
	
	Most galaxies with bulges harbor a supermassive black hole in their center \citep{Magorrian:1998p9015}, yet only a small fraction exhibit the powerful radiative or kinetic output associated with an active galactic nucleus (AGN).  While it is well established that matter falling onto the supermassive black hole is emitted as energy, the source of this material remains highly controversial.  To understand what activates and continues to fuel AGN, we must better characterize the conditions of the host galaxies in which they are found.  What environmental factors trigger these black holes to begin emitting so much energy and what continues to fuel this process? 
	
	Numerical simulations suggest that quasars (L$_{\mathrm{bol}}$$>$$10^{45}$ erg s$^{-1}$) are the end product of mergers between gas-rich disk galaxies, and that supermassive black hole accretion heats the interstellar material and quenches star formation leading to passive elliptical galaxies \citep{DiMatteo:2005p5934}.  Alternatively, other simulations suggest sources other than mergers may fuel lower luminosity AGN, such as gas streaming down galactic bars or steady cold gas streams \citep{Mulchaey:1997p9021,Hopkins:2006p9574,Dekel:2009p9116}. 

	A number of observational studies have provided interesting yet contradictory results about the relationship between the host galaxy and the AGN.  A study of the host galaxies of X-ray selected AGN from the Extended Chandra Deep Field-South found that AGN are in the most luminous galaxies, with intermediate optical colors, and bulge dominated morphologies \citep{Silverman:2008p8462}.  Another study of narrow emission line (NL) AGN in the Sloan Digital Sky Survey (SDSS) found the hosts were predominantly massive early-type galaxies and the most luminous AGN galaxies had significant star formation \citep{Kauffmann:2003p2397}.  An additional survey of the SDSS NL AGN host galaxies found, compared to a sample of nearby inactive galaxies, most AGN occur along the red sequence  \citep{Westoby:2007p8467}.  Even though these studies draw their conclusions from large optical surveys or soft X-ray surveys, their results may be biased by missing an important population of obscured AGN.  

	A less biased survey of AGN must account for absorption of light by gas and dust.  In addition, the survey must distinguish between host galaxy emission primarily from stars and emission from AGN.  Historically, an inability to account for these factors provided different results depending upon the wavelength used for observation. Initially, AGN were selected by radio techniques \citep{Heckman:1980p9959} or optically using broad and strong emission lines, a luminous point-like nucleus, or irregular galaxy color \citep{Weedman:1977p8832}.  The presence of broad optical emission lines was used to separate Seyfert 1 galaxies with broad permitted and narrow forbidden lines and Seyfert 2 galaxies with only narrow permitted and forbidden line emission.  In the unified model (Antonucci 1993) both broad line Seyfert 1 galaxies and obscured Seyfert 2 galaxies are intrinsically the same, with the differences being the viewing of the central engine.  Surveys of AGN taken in the optical, UV, and soft X-rays ($<$ 2 keV) miss an important population of obscured narrow line AGN only visible in the ultra hard X-ray and mid-IR wavelengths \citep{Mushotzky:2004p8576,Koss:2011p12483}.  While the mid-IR wavelength is less obscured, this wavelength range is problematic because of confusion with emission from star formation, sensitivity to the amount of obscuring material, and the lack of a unique way to select AGN from other luminous IR galaxies \citep{Stern:2005p8982,Hickox:2009p707}.  Therefore, the ultra hard X-ray, $>$15 keV range offers an important new way to select AGN for a less biased survey. 

The BAT survey is an all sky survey in the ultra hard X-ray range that has identified 461 objects of which 262 are AGN \citep{Tueller:2010p6018}.  The BAT instrument is a large field of view (1.4 steradian) coded aperture imaging instrument.  Because of the large position error of BAT ($\approx2\arcmin$) higher angular resolution X-ray data for every source from Swift-XRT or archival data have been obtained allowing associations with 97\% of BAT sources.  This sample is particularly powerful since the BAT is sensitive in the 14--195 keV band and at obscuring columns of $>$$10^{24}$ cm$^{-2}$ where only high-energy X-ray emission (tens of keV) can pass through the obscuring material.  It is therefore sensitive to heavily obscured objects where even hard X-ray surveys (L$_{2-10 \: \mathrm{keV}}$) are severely reduced in sensitivity.    At 22 months\footnote{http://heasarc.gsfc.nasa.gov/docs/swift/results/bs22mon/}, the BAT survey has a sensitivity of approximately $2.2\times10^{-11}$ erg cm$^{-2}$ s$^{-1}$.  With this sensitivity, the BAT survey is about 10 times more sensitive than the previous all-sky ultra hard X-ray survey, HEAO 1 \citep{Levine:1984p9096}.   About 15$\%$, or 30 of the AGN, have never before been detected as AGN at other wavelengths.    

	While AGN host galaxy studies using X-ray surveys typically probe objects at moderate redshift (out to z$\approx$1), more nearby (z$<$0.05) AGN offer the best opportunity to study the host in detail since high spatial resolution data are easily obtainable. The majority of the BAT AGN are nearby with a median redshift of 0.03.  Thus, a study of the BAT AGN sample provides an excellent opportunity to answer the controversial question of AGN fueling and its relationship to the host galaxy.
	
	%The multiwavelength properties of the BAT AGN sample have been studied in a number of papers ().  The BAT AGN sample host galaxies have been studied in optical imaging by Schawinski et al.~2009 and Vasudevan et al.~2009.  However, these surveys were limited to 16 and 17 BAT AGN whereas we focus on a much larger sample of 185 BAT AGN.  In addition, these papers reached opposite conclusions, the Schawinski paper found AGN seeming to be in redder galaxies indicating the AGN may be suppressing star formation and the Vasudevan study found AGN in bluer galaxies.  With 10 times more AGN than the previous optical imaging studies, we have a much larger sample to study the host galaxies.
	
	The BAT AGN sample have already provided several interesting results about their host galaxies.   One study found that BAT AGN show a merger rate of 24\% compared to only 1\% in normal galaxies  \citep{Koss:2010p7366}.  A study of the morphologies based on NED classifications and DSS imaging found a majority to be in spirals or peculiars \citep{Winter:2009p79}.  Other studies found that BAT AGN have been shown to have additional reddening of the narrow line region not accounted for in optical studies and be misclassified as star forming or composite regions \citep{Winter:2010p6825,Melendez:2008p3807}.  In the case of host galaxy colors, two studies using $<$20 BAT AGN reached different conclusions: \citet{Schawinski:2009p1181} found that the AGN tend to be in intermediate or red galaxies and possibly suppress star formation, while \citet{Vasudevan:2009p7223} found BAT AGN in blue, starforming galaxies.  This contradiction highlights a major problem in current AGN host galaxy studies: the paucity of large, uniform samples with high quality data.  To make progress on this issue, we have assembled the largest sample of ultra hard X-ray selected AGN with host galaxy optical data to date, with 185 AGN in total.   The BAT AGN and comparison samples are discussed in \S2; data reduction and analysis in \S3 with additional discussion of removal of AGN light in appendix \S A, the comparison sample in \S B,  and selection effects in appendix \S C; the results in \S4; and the summary and discussion in \S5.

\section{Samples}	
\subsection{BAT AGN Sample}

We selected our sample to focus on Seyfert AGN in the BAT sample which contribute the large majority, 87$\%$ of the total sample, and 99$\%$ of the nearby, z$<$0.05 sample.     We use the classifications of \citet{Tueller:2010p6018} to exclude beamed sources such as blazars.  We also restrict the sample to nearby AGN (z$<$0.05) which includes 80\% of the BAT sample.  These AGN can be further classified on the basis of broad or narrow emission lines.  We define NL AGN as having H$\alpha$$<$$2000$ km/sec  using spectroscopic data from \citet{Winter:2010p6825}, \citet{Ho:1997p5224}, and the SDSS.  For those BAT galaxies without spectra, we use the SDSS galaxy class separator or available data from NED to separate NL and broad-line objects.  The BAT sample has 50\% (93/185) NL and 50\% (92/185) broad-line sources.   In addition to including the entire range from unobscured to highly obscured, the AGN have moderate luminosities ($\log$ L$_{2-10\,\rm{keV}}$$\approx$43), and therefore accretion rates, typical of the local universe.

%These AGN hold the best possibility of understanding what properties of the host galaxy trigger the AGN.  This is because in a typical Seyfert galaxy, the energy emitted by the nuclear source at visible wavelengths is comparable to the total energy emitted by all the stars in the galaxy whereas in quasars the energy of the nuclear source is brighter by a factor of 100 or more, greatly diluting the host galaxy light.  Also, because quasars are extremely luminous they are predominantly found at large distances.  Current theory says AGN are all just different manifestations of the same "standard" phenomenon; a SMBH accreting gas onto a disk and blasting it back out into space in collimated jets.  In this model the differences come from obscuration, viewing angle, and accretion rate.  The difference between Seyfert 1 and Seyfert 2 AGN is due to obscuring material in the line of sight. 

  	By imposing an upper redshift limit (z$<$0.05) to the sample, we focus on galaxies that are close enough to have good spatial resolution (700 pc) from ground-based optical imaging.  We further restrict our survey to Northern declinations ($>$-25$\degree$).  We also exclude six nearby galaxies with bright foreground Galactic stars where photometry is difficult and three nearby bright galaxies with saturated pixels.    This survey covers 125 nearby AGN or 50$\%$ of the entire BAT AGN sample from the 22 month survey.  We also included an additional 60 AGN galaxies detected in the 58 month BAT sample catalog (Baumgartner et al.~2010, submitted) with X-ray followup with the $\Sw$ XRT telescope that showed a coincident source to the 14--195 keV BAT source in the 2--10 keV band.   

The optical imaging data for these galaxies include 17 nights at the Kitt Peak 2.1m telescope in the  \textit{ugriz} SDSS bands (Table~\ref{allagnkp}) and data from the SDSS (Table~\ref{allagnsdss}).  The final Kitt Peak and SDSS sample includes a total of 185 galaxies, 79 BAT AGN host galaxies observed at Kitt Peak, 92 from the SDSS, and 14 galaxies observed by both the SDSS and at Kitt Peak.

\subsection{Comparison Samples}
	To better understand the host galaxy properties of BAT AGN, we used a comparison sample of inactive galaxies and a sample of emission line selected AGN from the SDSS. We will henceforth refer to the three samples as the \emph{BAT AGN, inactive galaxies}, and \emph{SDSS AGN}, respectively.
	
	The inactive galaxies were selected from the SDSS to have high quality photometry and similar redshifts as the BAT AGN.  We selected all non-QSO galaxies from the SDSS DR7 with spectra and imaging data with redshift confidence, z$_{conf}$$>$0.9 and a redshift interval similar to the BAT AGN (0.01$<$z$<$0.07).  We chose this slightly higher redshift interval because many of the SDSS galaxies with z$<$0.01 are too bright to be targeted in spectroscopy.  We also removed NL Seyfert or LINER AGN from this sample using emission line diagnostics \citep{Kewley:2006p1554} and the Garching catalog of reduced spectra of narrow line AGN \citep{Kauffmann:2003p2397}.  Galaxies totaled 68,275.  We will refer to this sample as the \emph{inactive galaxy sample}.

	Finally, we used a sample of emission line selected AGN in the SDSS for comparison, which we refer to as the \emph{SDSS AGN}. \citet{Winter:2010p6825} found that the majority (75\%) of a sample of 64 BAT AGNs were Seyferts.  Only 3/64 ($<$5\%) of BAT AGN were classified as LINERS, so we excluded this type of AGN from the comparison sample.  We chose narrow-line AGN since the nucleus is invisible in the optical band and thus does not have to be modeled to determine the host galaxy properties.  We therefore chose a sample of all type 2 Seyferts in the SDSS DR7 with 0.01$<$z$<$0.07.    We used 1282 Seyferts in this redshift range.  
	
\begin{figure} 
\plotone{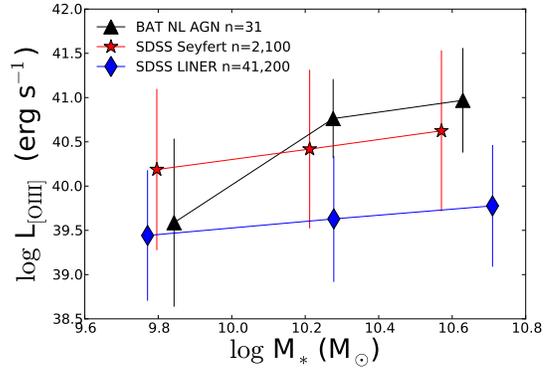}
\caption{Average [O \textsc{III}] luminosities for the NL AGN in the BAT sample compared to the SDSS sample of narrow-line Seyferts and LINERS.  The error bars indicate standard deviations in each bin.  We find that the distributions of [O \textsc{III}] from the NL BAT AGN and NL SDSS Seyferts used in this study are from the same parent population (P(K-S)=82\%).  The [O \textsc{III}] luminosity is a measure of AGN power and suggests the NL BAT AGN have similar AGN power as SDSS NL Seyferts.  }
\label{o3comp}
\end{figure}

	To ensure that the BAT AGN were not more intrinsically luminous than the SDSS AGN sample of Seyferts, we compared the [O \textsc{III}]  of the BAT AGN with available spectra to the SDSS AGN (Fig.~\ref{o3comp}).  When measuring [O \textsc{III}], we used the narrow Balmer line ratio (H$\alpha$/H$\beta$) to correct for extinction assuming an intrinsic ratio of 3.1 and the \citet{Cardelli:1989p1821} reddening curve.  For reference we also included LINERS in the SDSS.  For the BAT AGN we used spectroscopic data from \citet{Winter:2010p6825}, \citet{Ho:1997p5224}, and the Garching Catalog of SDSS spectra \citet{Kauffmann:2003p2397}.  We find that the BAT AGN have similar [O \textsc{III}] luminosities as the SDSS NL Seyferts, suggesting that they also have similar intrinsic luminosities (although there may still be differences, see \S5).  We find a similar relation when only including sources with SDSS spectroscopy and excluding spectroscopic data from \citet{Winter:2010p6825} and \citet{Ho:1997p5224}.
	
\begin{figure}[t!] 
\centering 
\plotone{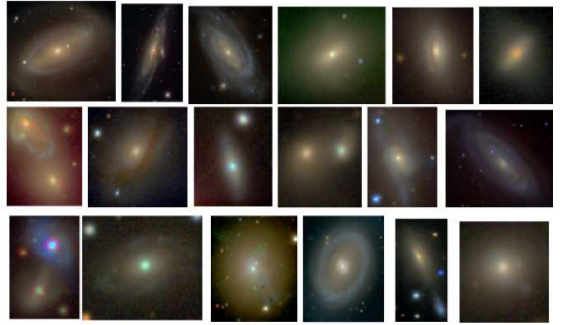} 
\caption{Random sample of 18  \textit{gri} composite images of galaxies in the BAT AGN sample taken at Kitt Peak. An arcsinh stretch was used with flux scaled by magnitudes \citep{Lupton:2004p7328}.  } 
\label{f3}
\end{figure}

\section{Data Acquisition, Reduction, and Analysis}

Throughout this work, we adopt the following cosmological parameters to determine distances: $\Omega_{\rm m}$= 0.27, $\Omega_{\rm \Lambda}$= 0.73, and $H_{\rm 0}$ = 71 km s$^{-1}$ Mpc$^{-1}$.  For galaxies with z$<$0.01, we use the mean value of redshift independent measurements from NASA Extragalactic Database (NED) when available.  Unless otherwise noted, error bars correspond to 1$\sigma$ standard deviation of the sample.

\subsection{Data Acquisition}
The Kitt Peak sample was obtained in February 2008 and November 2008 using the t1ka, t2ka, and t2kb CCDs on the 2.1m telescope.  A random sample of  \textit{gri} tricolor images that have been flux calibrated using the procedure of \citet{Lupton:2004p7328} are shown in Fig.~\ref{f3}.  Each galaxy was imaged so that a high signal-to-noise ratio could be achieved to capture faint features and low surface brightness emission with 12 minutes in \textit{u} and 6 minutes in \textit{griz}. The \textit{u} images were obtained at twice the imaging time because of the higher sky brightness and lower signal-to-noise ratio in this band.  Limiting magnitudes and observing conditions can be found in Table \ref{limmagcomp}.

\subsection{Initial Calibration}
	The initial imaging analysis involved calibration of zero point magnitudes, coadding SDSS plates for larger galaxies, and removing extraneous sources (e.g.~foreground stars, other galaxies). For calibration of the Kitt Peak data, we used the primary standard star network of 158 stars used by the SDSS for calibration \citep{Smith:2002p9117}.  A calibration star was imaged before and after each galaxy at similar airmass.  Extinction coefficients were determined using standard stars on a nightly basis.  Standard IRAF routines were used to remove bias, dark current, and CCD non-linearity.  Galactic extinction corrections were made based on data from \textit{IRAS} and \textit{COBE/DIRBE} \citep{Schlegel:1998p9121}.  For 8 SDSS galaxies extending across multiple SDSS plates, the MONTAGE software was used to reproject the images, rectify background and coadd plates.  WCSTools was used to register Kitt Peak images from USNO stars \citep{Mink:1996p9129}.  Nearby foreground stars and galaxies were identified using segmentation maps produced by SExtractor \citep{Bertin:1996p9137}.  The segmentation map identified object pixels using a threshold of 3$\sigma$ and a minimum of 5 pixels in \textit{griz}.  In the \textit{u} band, the detection threshold was set to 1$\sigma$ to ensure that faint star forming regions were detected.    The segmentation map was visually compared to three color  \textit{gri} images to ensure proper star and galaxy separation.  Stellar or foreground galaxy objects were masked with the IRAF FIXPIX routine.

%Finally, the Kitt Peak sample was 0.11$\pm$ mag brighter than the SDSS. This is consistent with photometric comparison in the SDSS by other groups that found an overestimation of sky brightness near bright galaxies (Lauer et al.~2007; Bernardi et al.~2007; Lisker et al.~2007). 	
%We used the  \textit{ugriz} filter curves of the SDSS for a point source, and then normalized based on the assumption that the filters are AB system with a magnitude 0 object having 3631 Janskys in each band.   The final AGN colors in u-g, g-r, r-I and i-z were [0.146,0.151,0.105,0.096].  We used the mean Seyfert 2 galaxy colors of [1.47,0.62,0.34,0.27] for our BAT sample host galaxies to find that comparable AGN\% of total light in other bands compared to the r band. So if the AGN \% of total light in the r band is 50\%, then the comparable \%AGN in the ugiz band would be [84, 61,45,41].

\subsection{Host Galaxy Photometry}
	
	The residual host galaxy colors were measured after removing the AGN light contribution using GALFIT.  To ensure that the AGN emission was properly removed we performed simulations of AGN galaxies to test GALFIT and also checked the subtraction for BAT AGN (see Appendix \S A for a discussion).  A modified form of the Petrosian system, the same as is used in the SDSS automated pipeline, was used for photometry \citep{Blanton:2001p8687} for both the SDSS and Kitt Peak observed BAT AGN galaxies.  The Petrosian aperture is determined to be large enough to enclose almost all of the flux for typical galaxy profiles, but small enough that the sky noise is not significant.    For consistency, the Petrosian aperture is determined from the $r$ band and applied to the other bands.   A galaxy with a bright stellar nucleus, such as a broad-line Seyfert galaxy, can have a Petrosian radius set by the nucleus alone; in this case, the Petrosian flux misses most of the extended light.  Therefore the Petrosian radius in the $r$ band was determined after the AGN model from GALFIT was subtracted.  We then used the software KCORRECT \citep{Blanton:2007p3139} with the \textit{ugriz} photometry to calculate the stellar masses.  This code uses the stellar population models of \citet{Bruzual:2003p1640} and photoionization models of \citet{Kewley:2001p5316}.  We used the same software to calculate the stellar mass for the inactive galaxy and SDSS AGN control samples. 
	
%To determine this aperture, the radius where the Petrosian Ratio is equal to 0.2 was determined with 
%$R_{\rm P}=(\int_{0.8r}^{1.25r} dr^{\prime}~2 \pi r^\prime I(r^\prime)/\pi(1.25^2-0.8^2) r^2)/(\int_{0}^{r^\prime} dr^\prime~2 \pi r I(r^\prime)/\pi r^2)$. 

	We also made a comparison of the overlapping galaxy data from the SDSS and Kitt Peak to ensure there were no systematic differences in photometry.   In some cases, the automated SDSS pipeline's photometry shreds bright galaxies into many smaller galaxies which leads to incorrect photometry estimates for bright, nearby galaxies (see Appendix \S B).  For cases where shredding wasn't a problem, the colors of  overlapping galaxies observed at Kitt Peak and in the SDSS showed good agreement  in \textit{griz} with mean color differences less than 0.02 mag and sample standard deviation less than 0.05 mag.  The \textit{u} band is more uncertain with mean $(u-g)$= $0.05\pm0.16$ mag brighter for the Kitt Peak measurements.  This difference is expected because the \textit{u} band measurements in the SDSS have a lower signal-to-noise ratio and also suffer from a red leak\footnote{http://www.sdss.org/dr7/algorithms/fluxcal.html}.  	
		
	 There were also ten nearby galaxies or 5\% of the sample with pixel saturation in the SDSS images because of a bright nucleus.  For these galaxies we masked the saturated pixels and fit the remaining image with a point source (PS) and floating S\'{e}rsic Index.  We then used the model fit to recover the saturated pixels.  We restricted using this method to those galaxies with saturation in a single imaging band with  $<$25 pixels saturated.  This excluded three very nearby galaxies (z$<$0.005; NGC 1068, NGC 4151, NGC 3998) with saturated pixels in multiple filters from the study.  For five images of nearby galaxies taken at Kitt Peak, the estimated Petrosian radius extended beyond the edge of the CCD.  In these cases we used the maximum part of the image available to determine the photometry.  In addition, 16 galaxies or 9\% had very low signal-to-noise ratio measurements in the \textit{u} band and were not included in the photometry.
	 
	 We have also provided a detailed discussion of the selection effects in the X-ray selected BAT sample of AGN, and the optically selected SDSS AGN and inactive galaxies (see Appendix \S B and C).  These selection effects are also included to enable comparison with AGN surveys at other wavelengths and for comparison with high-z studies.

\section{Results}

\subsection{Colors, Internal Extinction, and FIR emission}
\begin{figure*}[t] 
\centering
\plottwo{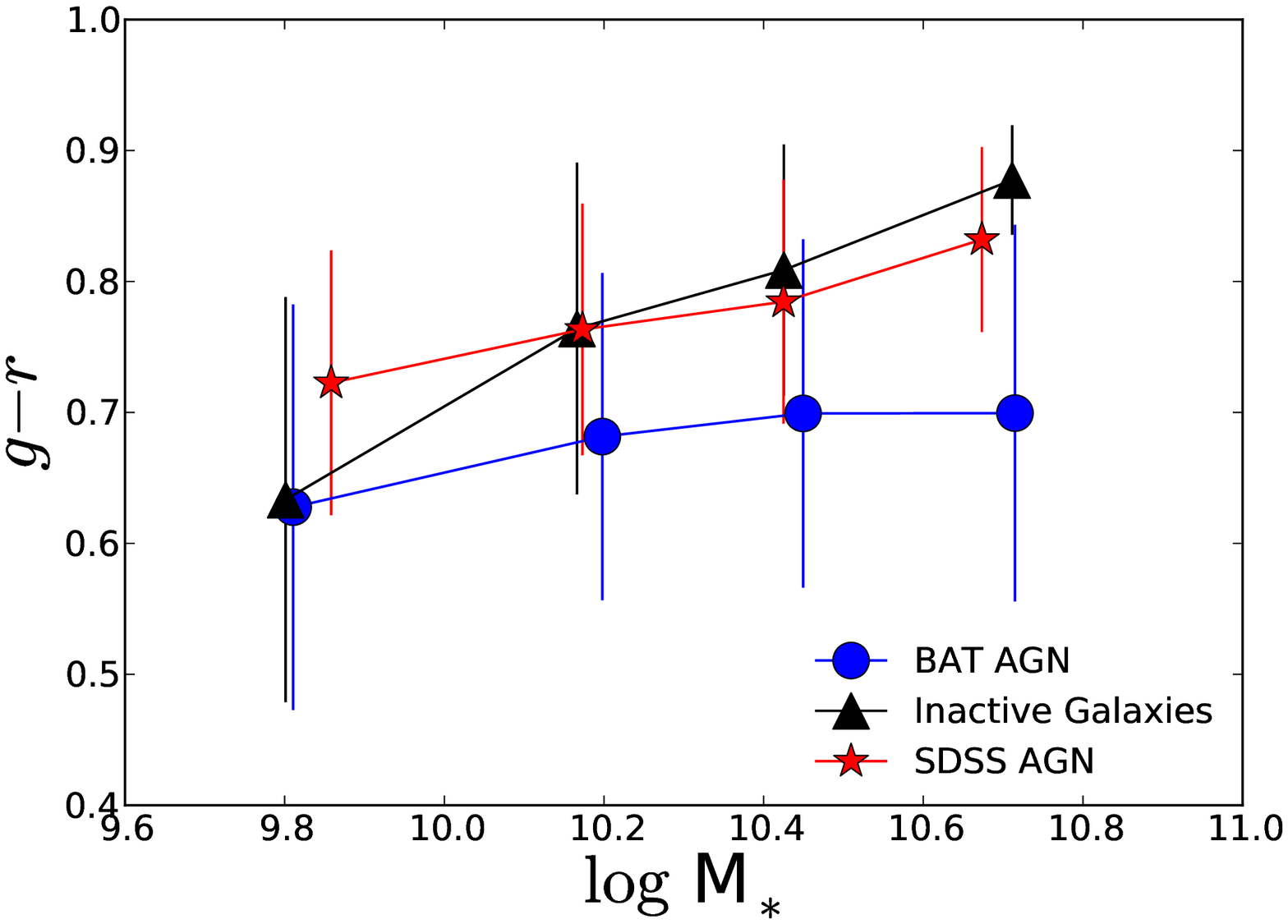}{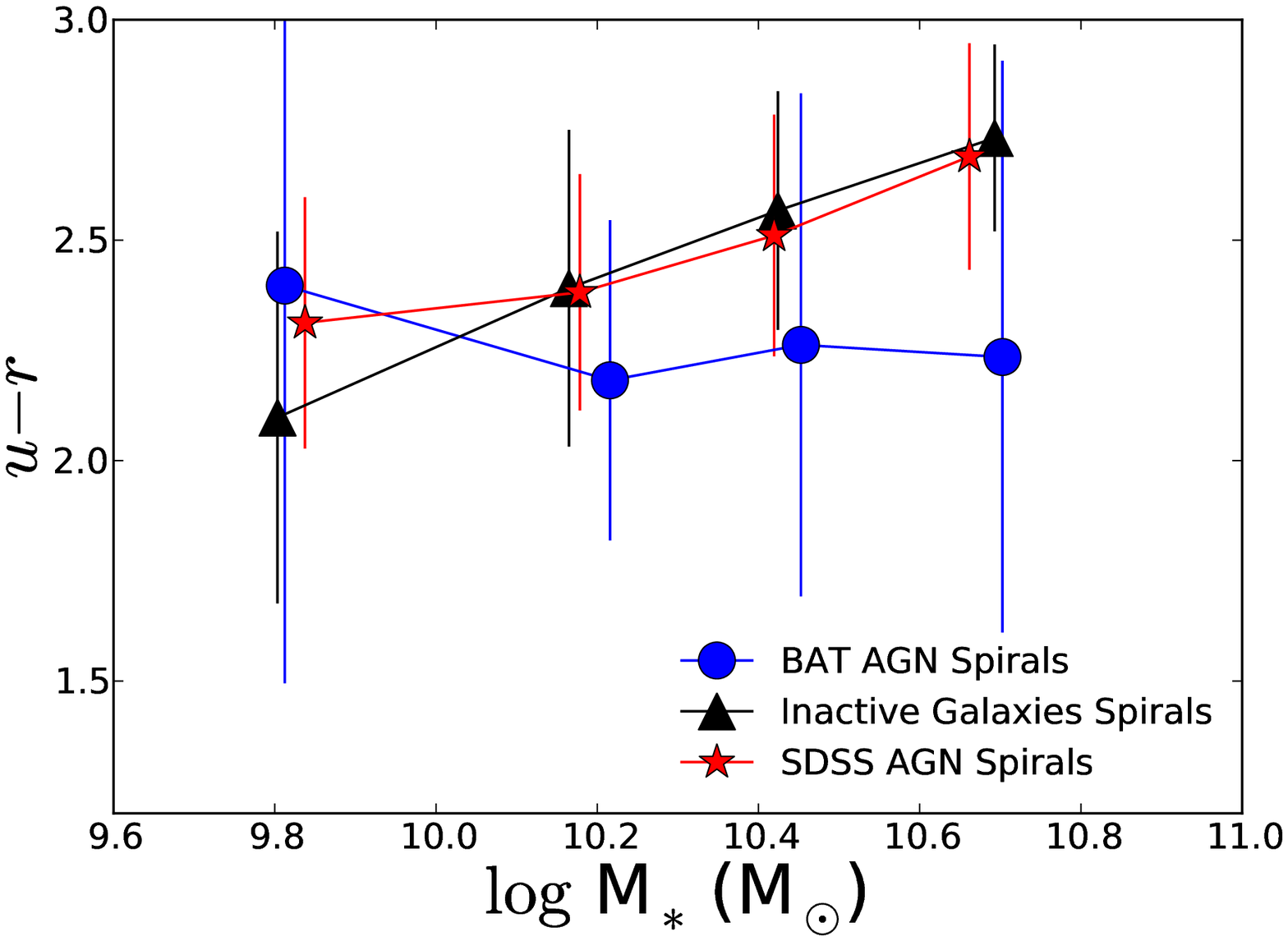}
\caption{Plot of average Petrosian $g-r$ and $u-r$ for BAT AGN, inactive galaxies, and SDSS AGN.  We find that the BAT AGN have bluer average colors than inactive galaxies or SDSS AGN in all but the lowest stellar mass bin.  The error bars indicate standard deviations in each bin.} 
%\vskip -.5 in
%\hskip 1.1 in 
\label{f31}
\end{figure*}

	Since host galaxy color traces star formation, a comparison of BAT AGN host galaxy colors to a sample of inactive galaxies should show whether the AGN is linked to enhanced or suppressed star formation.   A full listing of the photometry measurements of the different samples can be found in Table~\ref{tablephot}. A plot of $g-r$ and $u-r$ for BAT AGN, inactive galaxies, and optical AGN can be found in Fig.~\ref{f31}.  We find that the BAT AGN are bluer in both $g-r$ and $u-r$ than the sample of inactive galaxies and SDSS AGN in all but the lowest stellar mass bin (Table~\ref{tabcolor}). A Kolmogorov-Smirnov (K-S) test indicates a $<$1\% probability that the distribution of host galaxy colors for the BAT AGN are from the same parent distribution as the inactive galaxies or SDSS AGN.

\begin{figure} 
\centering 
\includegraphics[width=8.1cm]{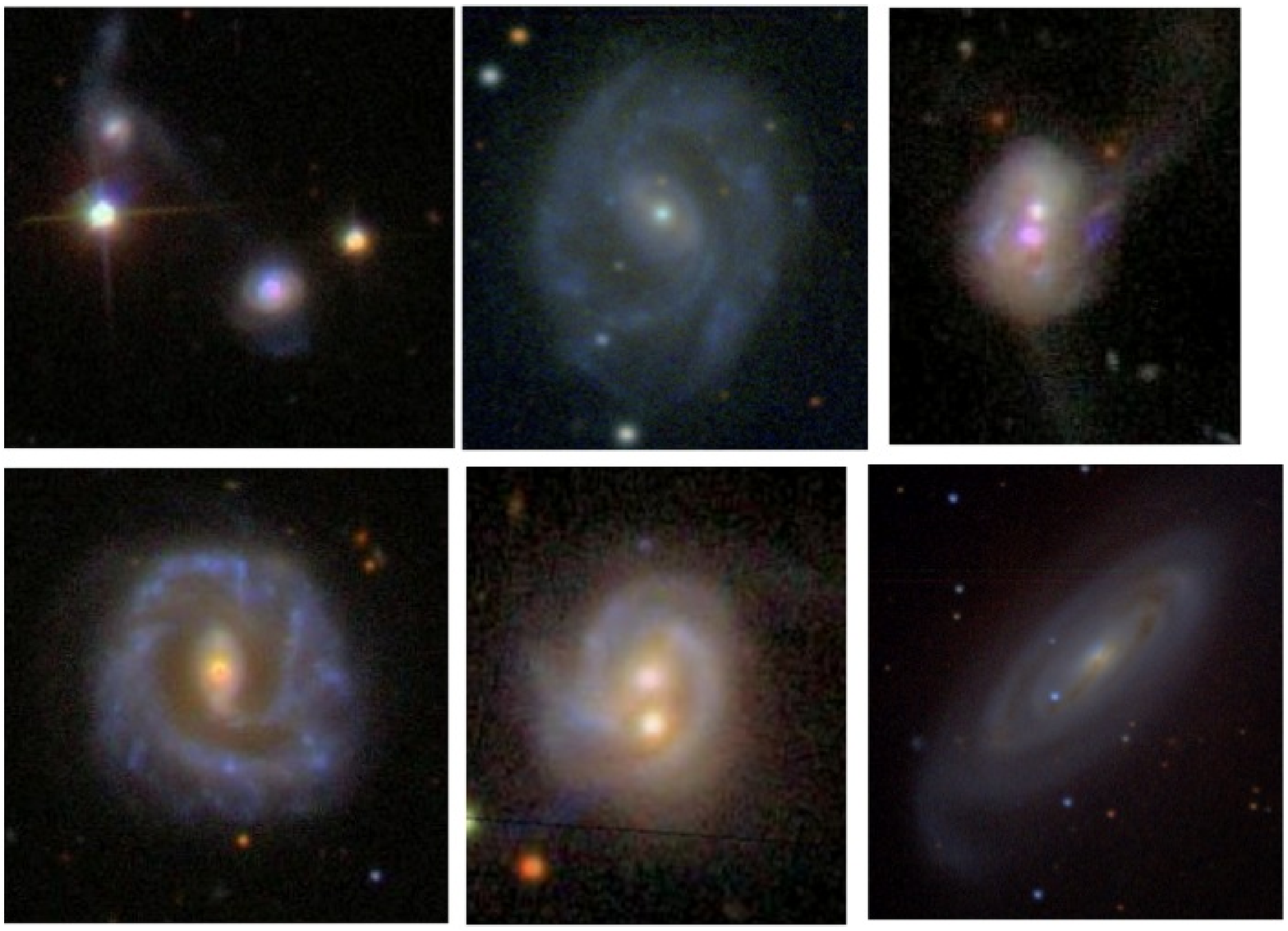}
\includegraphics[width=8.1cm]{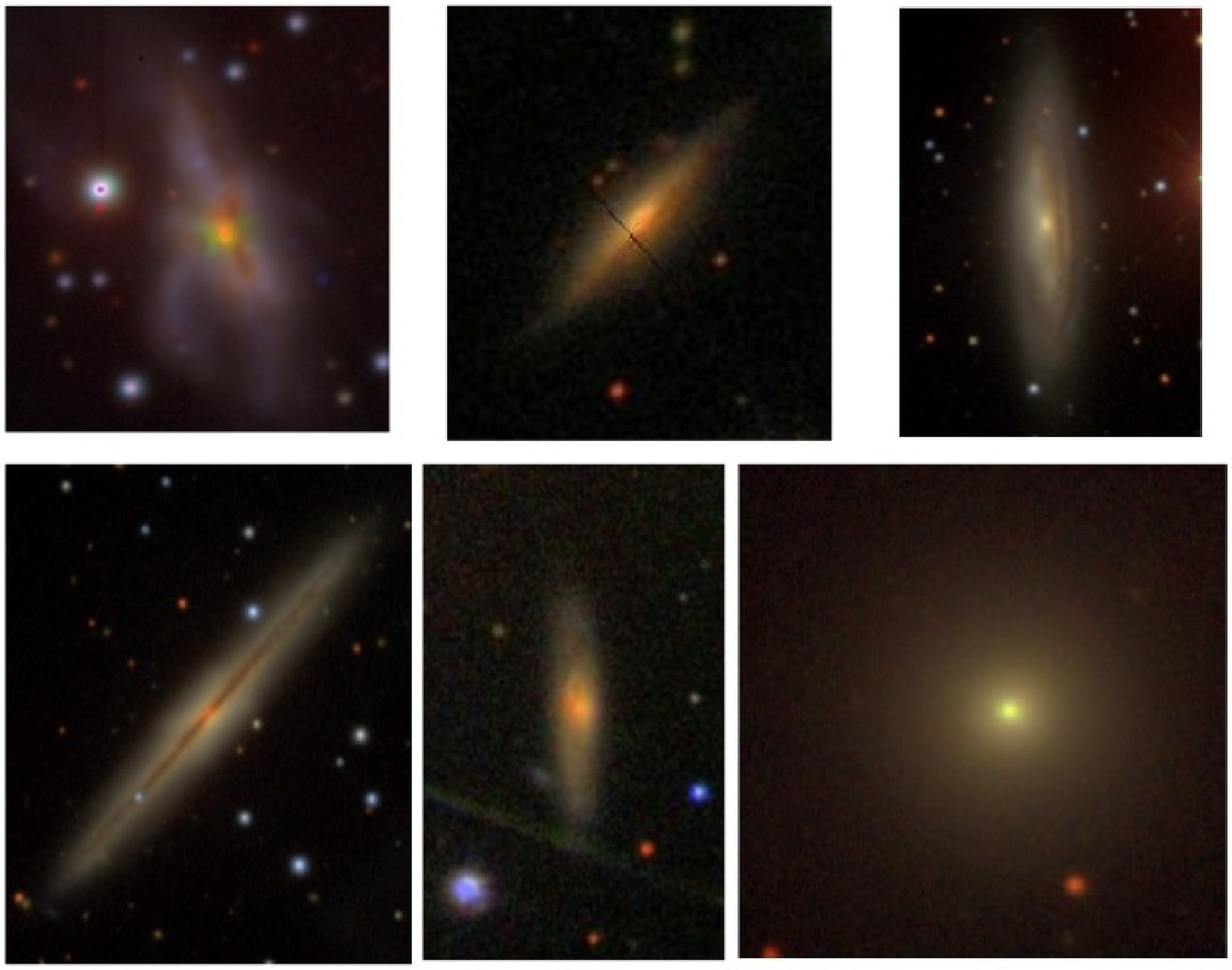}
\caption{{\em Upper}:  Tricolor images of six bluest BAT AGN host galaxies in $g-r$.  We find that these BAT AGN are predominantly in mergers and face on spirals.  {\em Lower}:  Tricolor images of six reddest BAT AGN host galaxies in $g-r$.  We find that 4/6 of these BAT AGN are in edge-on spirals that are reddened by internal extinction.  The predominance of face on spirals in the bluest galaxies and edge-on spirals in the reddest galaxies indicate the need for a reddening correction based on inclination.   } 
\label{extinctimage}
\end{figure}

	Host galaxy colors can be reddened in galaxies with high inclinations and large amounts of dust, so we measure whether these corrections change our result that the BAT AGN tend to be bluer than inactive galaxies or SDSS AGN.  Because of the relative uncertainty in these measurements, we do not apply individual reddening corrections on any plots, but estimate how reddening corrections effect the average colors of the samples.  In Fig.~\ref{extinctimage}, we show tricolor images of the 6 reddest and bluest BAT AGN host galaxies in $g-r$.   The predominance of face-on spirals in the bluest sample and edge-on spirals in the reddest sample indicates the need for reddening correction based on inclination.  \citet{Masters:2010p8988}  studied internal extinction of galaxies identified as spirals in Galaxy Zoo and provided corrections based on the measured inclination from the SDSS photometry.   For the three samples, we find that the average extinction for spiral galaxies is similar (0.04$\pm$0.03 in $g-r$), but because of the higher number of spirals in the BAT AGN sample compared to SDSS AGN or SDSS inactive galaxies (see \S4.2), there is a larger extinction correction for the BAT AGN sample. 
 
	There are galaxies that have high levels of star formation, but are reddened in the optical wavelengths because of the presence of dust.  For example, in the BAT AGN sample, the reddest galaxy in $g-r$ is NGC 6240, a luminous infrared galaxy (LIRG) in a major merger, which shows a large amount of star formation in the far-IR.  This indicates that dust can play a significant role in reddening the optical light and hiding increased levels of star formation.  Therefore, we estimate the total dust extinction ($A_V$) for each galaxy by fitting the host $ugriz$ SEDs using FAST \citep{Kriek:2009p9587} with single-burst stellar population models.  We use stellar templates from \citet{Bruzual:2003p1640} with the \citet{Chabrier:2003p9594} initial mass function and solar metallicity.  We fit a \citet{Calzetti:2000p9589} dust extinction reddening law (A$_V$= 0--3).  We assume that R$_V$=3.1, and therefore E(B-V)=A$_V$/3.1.  We find that the reddening determined using fits to the optical SED of BAT AGN (0.34$\pm$0.26 in $g-r$)  are on average larger than for comparable inactive galaxies (0.20$\pm$0.14) or SDSS AGN (0.21$\pm$0.13).  Therefore, we do not find evidence that the bluer measured colors of BAT AGN, compared with those of inactive galaxies and SDSS AGN, can be explained as smaller dust reddening.  In fact, reddening corrections make the BAT AGN even bluer than comparable SDSS AGN or inactive galaxies.  We also find that the reddening corrections for host galaxy inclination are on average much smaller than dust reddening (0.04$\pm$0.03 vs.~0.34$\pm$0.26 in $g-r$). 

	The far-IR provides an additional tracer of star formation that is less sensitive to reddening than shorter wavelengths.  The 90 $\mu$m emission is a useful tracer of strong bursts of recent star formation and is less affected by AGN emission \citep{Netzer:2007p1714,Mullaney:2011p12174}.   We first looked at the rate of detection in \textit{AKARI}  in each of the samples in the same redshift range (0.01$<$z$<$0.05).  In this redshift range, 54$\pm$5\% (86/185) of BAT AGN are detected by \textit{AKARI}  at 90 $\mu$m, compared to only 4$\pm$1\% of SDSS AGN, and 5$\pm$1\% of inactive galaxies.  The error bars represent 1$\sigma$ Poisson statistics.

	To ensure this difference was not an effect of the differences in the distribution of stellar mass and redshift of the samples, we matched each BAT AGN to one inactive galaxy and one SDSS AGN based on redshift and stellar mass.    We find a similar percentage for both inactive galaxies 7$\pm$3\% (11/158) and SDSS AGN 3$\pm$2\% (5/158) detected by \textit{AKARI}.  The error bars represent 1$\sigma$ Poisson statistics.
	
% and SDSS AGN 3\% (5/158) detected by \textit{AKARI}.  
	
	While there is a possibility of AGN contamination in the 90 $\mu$m emission, the similar levels of \textit{AKARI}  detections for the inactive galaxies and SDSS AGN indicate that this level of contamination is limited.  In addition,  the AGN contamination to the FIR should be similar for BAT and SDSS AGN since the [O \textsc{III}] luminosities are similar and  [O \textsc{III}] is an indicator of bolometric luminosity.  All of these results indicate that the BAT AGN are more luminous at 90 $\mu$m, which suggests enhanced star formation among BAT AGN when compared to SDSS AGN and inactive galaxies.

\subsection{Host Galaxy Morphology}

	We investigated galaxy morphology to find which environments are most conducive to hosting an AGN and how ultra hard X-ray selected AGN are different than the SDSS AGN or inactive galaxies.   A full listing of the morphological measurements can be found in Table~\ref{tabmorph}.  While we have limited our results to NL AGN because of the difficulty of subtracting the light distribution to make morphological measurements  \citep{Pierce:2010p8180}, we provide morphological measurements of broad-line AGN after AGN light subtraction with GALFIT in Table~\ref{tabmorph}.
	
\begin{figure} 
\centering 
\includegraphics[width=8.1cm]{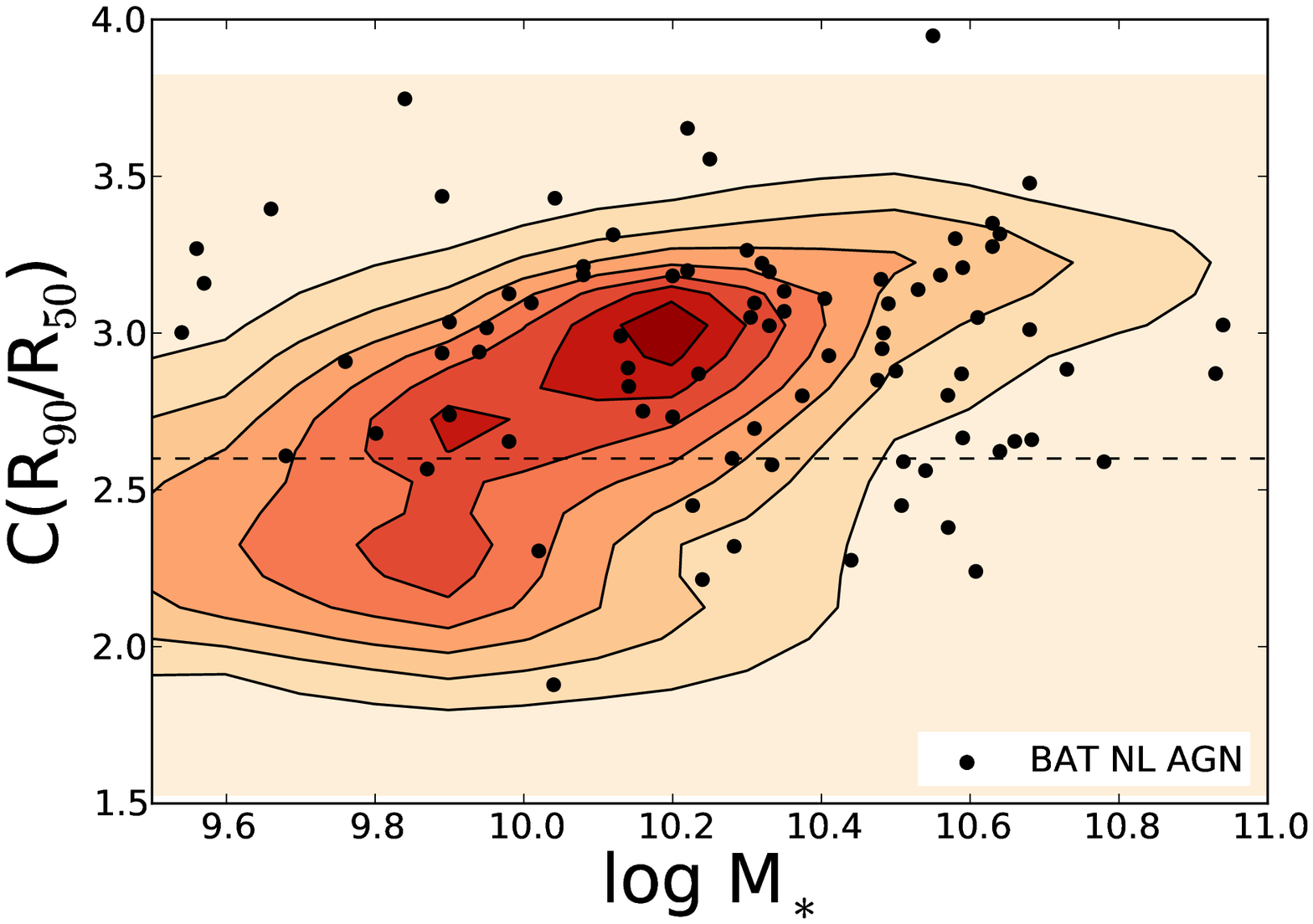}
\includegraphics[width=8.1cm]{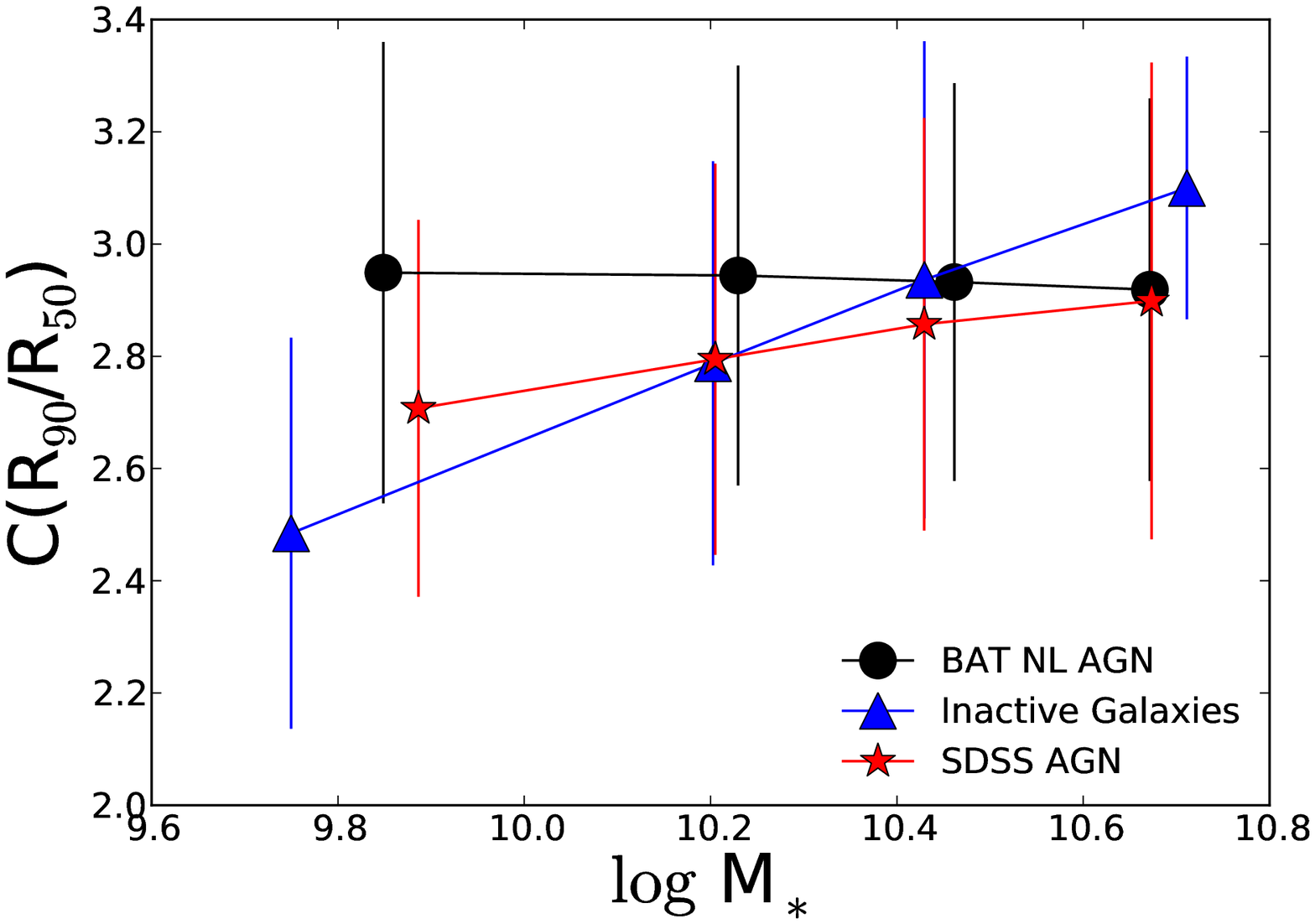} 
\caption{{\em Upper Panel}: Plot of concentration vs.~stellar mass with contours showing inactive galaxies and points for the BAT AGN. The contour levels are plotted on a linear scale with each level representing 10\% of the data (6875 inactive galaxies).  A dashed line is shown at C=2.6 at the separation between early and late types.  {\em Lower Panel}:  Plot of average concentration index by stellar mass.  We find that at low stellar mass ($\log$ M$_*<10$), BAT AGN tend to have higher concentrations than inactive galaxies or SDSS AGN.  The error bars indicate standard deviations in each bin.    } 
\label{conc}
\end{figure}

	The first measure we compared was concentration.  To enable comparison with the SDSS, the concentration index is defined as the ratio of the radii containing 90 and 50 per cent of the Petrosian $r$-band galaxy light $C=R_{\rm{90}}/R_{\rm{50}}$.  A galaxy with a steep concentration profile, such as an elliptical, will show a relatively large value for the $C$, while galaxies with a more shallow light profile, such as spiral and irregular galaxies, will have a lower $C$.  In addition, the concentration index is strongly correlated with the galaxy's bulge to total luminosity ratio as well as the supermassive black hole mass.  A plot of concentration vs. stellar mass is shown in Fig.~\ref{conc} for BAT AGN compared to inactive galaxies and SDSS AGN.  We find that at low stellar mass ($\log$ M$_*<10$), BAT AGN tend to have higher concentrations than inactive galaxies or SDSS AGN indicative of stronger bulges or a larger fraction of elliptical galaxies.  However, as shown below, the BAT AGN sample has a very low elliptical galaxy fraction.  
	
	While $C$ measurements are useful, they have been shown to be more closely related to luminosity than morphology \citep{Gavazzi:2000p9275}.   Since the human eye has consistently proven better than computational techniques at identifying faint spiral structure in images \citep{Lintott:2008p9199}, we used a catalog of visual classifications from the Galaxy Zoo project DR1 \citep{Lintott:2008p9199}.   Before morphological classification, we matched each NL BAT AGN to one inactive galaxy and one SDSS AGN by redshift and stellar mass.  We then used the visual classifications of morphology to divide host galaxies into elliptical, spiral, intermediate, and peculiar/merger.  Each galaxy had on average 37 independent classifications for a total of over 15,000 classifications for the 3 samples.  Elliptical or spiral galaxies were defined as galaxies in which on average $>$80\% people selected this type.  We define the peculiar/merger category following \citet{Patton:2008p5024} and \citet{Koss:2010p7366} by requiring a projected separation of $<$30 kpc and a radial velocity differences of less than 500 km/s between the sample galaxy and its possible companion.  The remainder of galaxies we classify as intermediate.

%A random sample of these different galaxy types can be found in Fig.~\ref{morphsample}.
	
\begin{figure} 
\centering 
\plotone{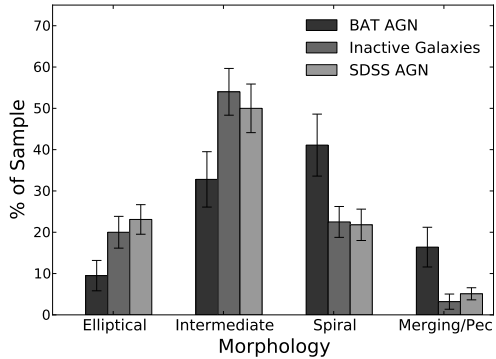} 
\caption{Histogram of morphologies from the BAT AGN, inactive galaxies, and SDSS AGN.  The morphologies were taken from measurements in the Galaxy Zoo DR1.  The error bars represent 1$\sigma$ Poisson statistics.  We find a higher incidence of spirals in the BAT AGN sample and less ellipticals and intermediates.  We also find more merging/peculiar types in the BAT AGN sample.} 
\label{zoo}
\end{figure}

	A comparison between the Galaxy Zoo classifications of the samples can be found in Fig.~\ref{zoo}.  We find that BAT AGN are more likely to be found in spiral morphologies at a rate (41\%) roughly twice that of inactive galaxies (22\%) or SDSS AGN (21\%).  We also find fewer BAT AGN in elliptical or intermediate types.  We confirm that BAT AGN are more likely to be found in merging systems consistent with the result of \citet{Koss:2010p7366}.  We see no statistically significant differences in the morphologies of inactive galaxies or SDSS AGN.  
	
	In addition, we looked at the Hubble Types of the BAT AGN sample compared to the Third Reference Catalog of Bright Galaxies (RC3) to confirm our results with Galaxy Zoo.  We used all galaxies in the RC3 in the same redshift range as the BAT AGN.  The RC3 is composed of bright galaxies with optical $B$ mag$<$15.5 and size larger than 1$\arcmin$.  This restriction excludes many faint galaxies or about 98\% of the SDSS sample in the same redshift range.  The BAT AGN sample has slightly higher optical luminosities than the RC3 (mean M$_{B}=-20.33\pm$0.82 vs. M$_{B}=-20.03\pm$1.03) and is at similar distances (mean z=0.025$\pm$0.01 vs. z=0.019$\pm$0.01 for the RC3).  We find more BAT AGN in early type Sa-Sb spirals (40\%) compared to the RC3 catalog (26\%).  We see fewer BAT AGN in ellipticals (3\%) compared the RC3 catalog (16\%).  The larger number of spiral morphologies in the BAT AGN sample is consistent with our analysis of morphologies using Galaxy Zoo.  
	
\begin{figure} 
\centering 
\plotone{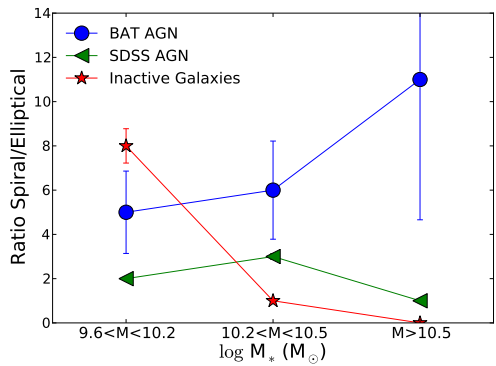}
\caption{Ratio of the number of spiral to elliptical galaxy morphologies as a function of stellar mass (M$_*$).  Galaxy morphologies are taken from Galaxy Zoo.   The error bars represent 1$\sigma$ Poisson statistics.  We find that the BAT AGN have a much larger number of massive spirals (log M$_*>$10.5) and very few ellipticals compared to inactive galaxies or optical AGN.} 
\label{massivespiral}
\end{figure}

	Since spiral galaxies tend to be found in less massive systems than elliptical galaxies, we examined the relationship between morphology and stellar mass.  For the BAT AGN in spirals, we find a higher average stellar mass ($\log$ M$_*$=10.34$\pm$0.27) than in ellipticals and intermediates ($\log$ M$_*$=10.07$\pm$0.42).  A K-S test indicates a $<$4\% the populations are the same.  This finding is in agreement with \citet{Schawinski:2010p6049} that found that optical AGN in elliptical systems tend to be in less massive systems.  We further investigated the predominance of massive spirals amongst BAT AGN by plotting the ratio of the number of spiral to elliptical galaxies by stellar mass (Fig.~\ref{massivespiral}).   For massive systems ($\log$ $M_*>$10.5), we find that BAT AGN are found in spirals at a rate that is 5 to 10 times higher than optical AGN or inactive galaxies.

\begin{figure} 
\centering 
\plotone{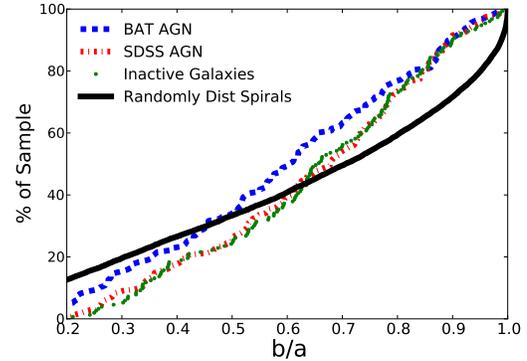}
\caption{Cumulative distribution of galaxy axis ratios in the $r$ band (b/a) for the NL BAT AGN, inactive galaxies, and SDSS AGN.  The axis ratio (b/a) is determined from the major and minor axes derived from SDSS isophotal photometry.  Randomly distributed spirals with a $\cos \theta$ distribution of axis ratios is shown for reference.  We find that more NL BAT AGN have b/a$<$0.55 than inactive galaxies and SDSS AGN. This limit is where \citet{Maller:2009p9276} found almost 90\% disk-like systems, and is consistent with increased incidence of spirals in NL BAT AGN compared to inactive galaxies and SDSS AGN.} 
\label{ba}
\end{figure}

	We also examined the galaxy inclination.  A study of SDSS galaxies by \citet{Maller:2009p9276} found that elliptical galaxies rarely have small axis ratios, and galaxies with b/a$<$0.55 are 90\% disk galaxies.  Therefore the axis ratio can be a reliable quantitative tracer of morphology.  To enable comparison with the SDSS catalog, we used the galaxy axis ratio in the $g$ band (b/a).  The axis ratio (b/a) is determined from the major and minor axes derived from SDSS isophotal photometry.  A plot of axis ratios of the samples can be found in Fig.~\ref{ba}.  For the axis ratios, we find a larger percentage of NL BAT AGN have b/a$<$0.55 which is where \citet{Maller:2009p9276} found almost 90\% disk-like systems.  This result is consistent with the increased incidence of spirals found by Galaxy Zoo for the BAT AGN.  We find no difference between the axis ratios of the inactive galaxies or SDSS AGN.  

	Previous studies have suggested that optical emission line classification of Seyfert galaxies may be missing a population of edge on galaxies \citep{Kirhakos:1990p9311,Simcoe:1997p9312}.  Since the NL BAT AGN are more likely to be in disk galaxies which have lower axis ratios, we separated the samples by morphology in Galaxy Zoo and then did a comparison of axis ratio.  In this case we do not see any statistically significant difference in axis ratios between NL BAT AGN spirals compared to spirals in inactive galaxies or SDSS AGN.
	  	  
	 Finally, the SDSS catalog provides independent measurements of the fraction of early type galaxies from the photometry and spectroscopy.  The SDSS spectroscopic parameter eClass classifies the spectral type of the galaxy using the principal component analysis technique, and the photometric parameter fracDev$_{r}$ measures the fraction of galaxy light that is fitted by a de Vaucouleurs law.  Following \citet{Bernardi:2003p9326} we define as early type galaxies those objects with eClass$<$0 from spectroscopy and fracDev$_{r}> 0.8$ from photometry.  In this comparison we only use NL BAT AGN with spectroscopy in the SDSS (185).  We find a statistically smaller number of NL BAT AGN in early type galaxies (39\%$\pm$8), compared to inactive galaxies (61\%$\pm$6) or SDSS AGN (56\%$\pm$6).  The error bars represent 1$\sigma$ Poisson statistics.  These results are consistent with a spectroscopic study of 64 BAT AGN which found that the majority of NL BAT AGN have spectra consistent with late type galaxies based on measurements of the stellar absorption indices \citep{Winter:2010p6825}.	  The percentage of BAT AGN galaxies classified as early type galaxies (39\%$\pm$8) using SDSS spectroscopy is significantly larger than that based on morphological measurements from the RC3 or galaxy zoo (3\% and 10\% respectively).  This is likely because the SDSS spectroscopy uses a 3\arcsec~aperture and is measuring only the central bulge portion of the galaxy.

	%We investigated the colors 16 overlapping BAT AGN galaxies published in Vasudevan et al. 2009 using UVOT data.  We find that for these overlapping galaxies the colors are an average of 0.2 mag redder in g  - r and 0.3 mag dimmer in the r band.  In addition,  4/16 were above the cutoff of 35\% of AGN light in the r band that indicated significant contamination of the KPNO and SDSS data.  The resolution of the UVOT is worse, ~2'' vs. 1.3''.  This suggests the bluer and brighter galaxy colors found using UVOT may be unidentified AGN contamination after using GALFIT.

\subsection{Colors, Morphology, and Ultra Hard X-ray Emission}
\begin{figure} 
\centering 
\plotone{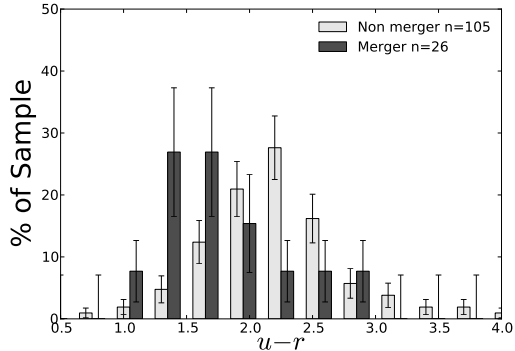}
\caption{Histogram of  $u-r$ colors of BAT AGN in major mergers vs.~BAT AGN that are not in major mergers.  The error bars represent 1$\sigma$ Poisson statistics. The merging galaxies show a bluer distribution of colors than non-mergers.  We find a similar trend in the $g-r$ colors. } 
\label{merg}
\end{figure}

\begin{figure} 
\centering 
\plotone{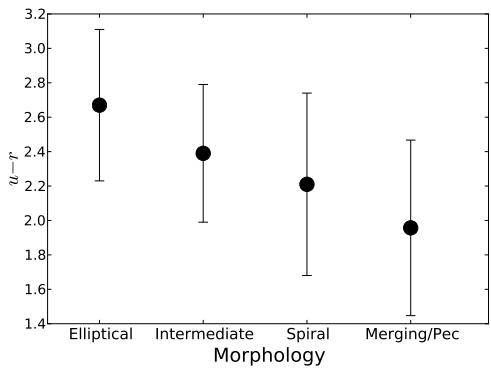}
\caption{Average $u-r$ color for different morphologies of BAT AGN.  The error bars indicate standard deviations in color for each morphology.   Galaxy morphologies were taken from Galaxy Zoo.  We find spiral and merger morphologies are on average bluer than elliptical or intermediate morphologies. We find a similar trend in the $g-r$ colors.        } 
\label{morphcolor}
\end{figure}

	Since we found BAT AGN host galaxies have a greater number of merger and spiral morphologies, an additional question is how this is related to host galaxy colors.  To test this we used a sample of BAT AGN in the process of mergers \citep{Koss:2010p7366} and did a comparison of their host galaxy colors compared to those BAT AGN not in mergers.  A histogram of the merger and non-merger sample is shown in Fig.~\ref{merg}.  When separated by color, we find the merging population showing bluer colors and hence increased levels of star formation.  This finding is in agreement with \citet{Koss:2010p7366} which found a higher rate of specific star formation from the \textit{IRAS} 60 $\mu$m fluxes for the merging BAT AGN.   We also find that spiral morphologies have bluer average colors than elliptical or intermediate morphologies (Fig.~\ref{morphcolor}). 

\begin{figure} 
\centering 
\plotone{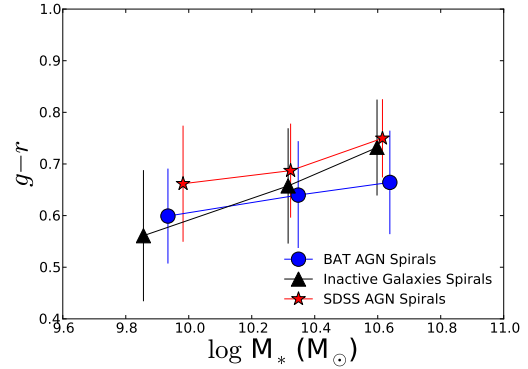} 
\caption{Plot of Petrosian $g-r$  for BAT AGN spirals, inactive galaxy spirals, and SDSS AGN spirals.  The error bars indicate standard deviations in each bin.  We find a much smaller difference in colors of the three samples when we compare only galaxies with spiral morphologies.  We find a similar trend in the $u-r$ colors.   This suggests that the higher incidence of spirals in the BAT AGN contributes to the bluer colors found when compared to the SDSS AGN or inactive galaxies. 
} 
\label{colorspiral}
\end{figure}

	Since galaxies of different morphologies tend to have different colors, we examined the colors of spirals in the BAT AGN, inactive galaxy, and SDSS AGN samples to look for differences.  In Fig.~\ref{colorspiral}, we show a plot of the colors of galaxies classified as spirals in Galaxy Zoo.  We find a much smaller difference in colors of the three samples when we compare only galaxies with spiral morphologies.  This suggests that the higher incidence of spirals in the BAT AGN sample may largely account for the bluer host galaxy colors when compared to SDSS AGN or inactive galaxies.
	
\begin{figure} 
\centering 
\plotone{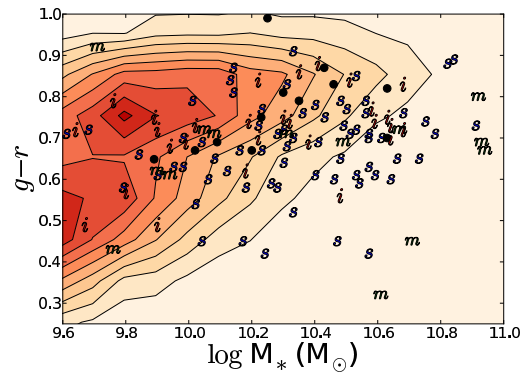}
\caption{Petrosian $g-r$ of BAT AGN (black dots and letters) and inactive galaxies (contours) plotted as a function of logarithm of the stellar mass (M$_*$).  The contour levels are plotted on a linear scale with each level representing 10\% of the data (6875 inactive galaxies).   In the BAT AGN sample, $m$, denotes a galaxy in a merger, $s$, a spiral morphology, $i$ an intermediate morphology, and black dots denote ellipticals.  We find a predominance of BAT AGN in blue, massive spirals and mergers in the regions outside of where most inactive galaxies lie.  We find a similar trend in the $u-r$ colors.} 
\label{contourmorph}
\end{figure}

	 A plot showing the color of each BAT AGN and its morphology can be found in Fig.~\ref{contourmorph}.  We also show the inactive galaxy colors with contours.  We find that the BAT AGN occupy a unique space in color, morphology, and stellar mass by tending to be in massive spirals and mergers that are bluer than massive ellipticals.  

	In terms of ultra hard X-ray luminosity, we do not find a significant difference between ellipticals, intermediates, or spirals.  However, we do find a larger mean ultra hard X-ray emission from BAT AGN in mergers ($\log$ L$_{14-195\,\rm{keV}}=$43.64$\pm$0.48) when compared to the non-merger sample ($\log$ L$_{14-195\,\rm{keV}}=$43.32$\pm$0.61.  A K-S test indicates a $<$5\% probability that the ultra hard X-ray emission from AGN in mergers is from the same population as the non-merger sample.

\subsection{Stellar Masses and Ultra Hard X-ray Emission}

\begin{figure} 
\centering 
\plotone{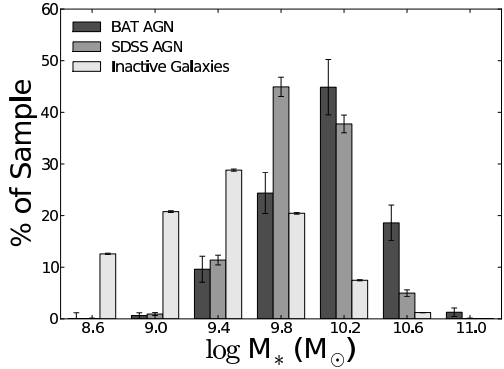} 
\caption{Histogram of stellar masses (M$_{*}$) of the BAT AGN sample compared to inactive galaxies and SDSS AGN.  The error bars represent 1$\sigma$ Poisson statistics.  The BAT AGN have significantly higher average stellar masses (mean $\log$ M$_{*}$=10.27$\pm$0.4) than inactive galaxies (9.45$\pm$0.58) and slightly higher stellar average stellar masses than SDSS AGN (10.18$\pm$0.28).} 
\label{histstellmass}
\end{figure}

	We find that the BAT AGN host galaxies are predominantly in the most luminous and massive of galaxies.  The mean optical luminosity is higher for BAT AGN (M$_{r}$ of -21.41$\pm$0.82) compared to inactive galaxies (-19.84$\pm$1.03),  and SDSS AGN (-20.95$\pm$0.69).  The BAT AGN also have higher mean stellar mass ($\log$ M$_{*}$=10.28$\pm$0.4) compared to inactive galaxies (9.46$\pm$0.58) and SDSS AGN (10.18$\pm$0.28).  This suggests that the BAT AGN tend to be in more massive galaxies than the SDSS AGN or inactive galaxies.  See Fig.~\ref{histstellmass} for a histogram of the stellar masses of the populations. A K-S test has $<$0.01\% probability that the BAT AGN stellar masses are from the same population as the inactive galaxies or SDSS AGN.  We also confirm that SDSS AGN are in more massive galaxies than inactive galaxies \citep{Kauffmann:2003p2397}.  We also fit a Schechter function and find that the logarithm of the characteristic stellar mass ($M^*$) from the best fit is 10.28, 10.02, and 9.89 for the BAT AGN, SDSS AGN, and inactive galaxies in agreement with our findings that BAT AGN are more massive than inactive galaxies or SDSS AGN.

%In Fig.~\ref{lumfunc}, we show a plot of the cumulative galaxy stellar mass function for the BAT AGN,  SDSS AGN, and inactive galaxy sample.  A single Schechter function has been fit to each sample.  We find that the logarithm of the characteristic stellar mass from the best fit Schechter function is 10.28, 10.02, and 9.89 for the BAT AGN, SDSS AGN, and inactive galaxies respectively.  
	
\begin{figure} 
\centering 
\plotone{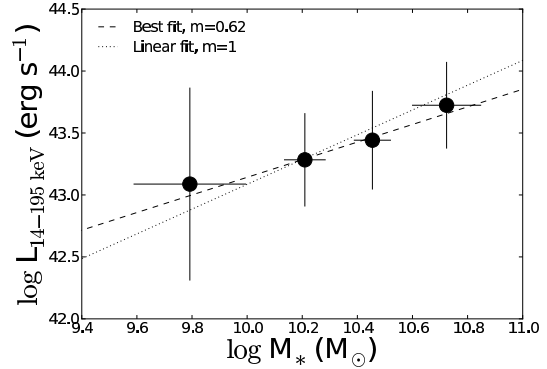} 
\caption{Plot of average ultra hard X-ray luminosity vs.~stellar mass.  Error bars represent the standard deviations in each bin.  We find greater ultra hard X-ray emission for galaxies with larger stellar mass.  A K-S test has $<$0.2\% probability that the distributions of ultra hard X-ray luminosities are the same for the lowest quartile and highest quartile of stellar mass.  In addition, for the average stellar masse, we find a correlation between $\log$ L$_{14-195\,\rm{keV}}$ and $\log$ M$_*$ with a slope of 0.62$\pm$0.17 and a less than 2\% probability that the values are uncorrelated.  We have also plotted a line with a linear fit between stellar mass and ultra hard X-ray emission.
 } 
\label{massbat}
\end{figure}

	We also find that the average hard X-ray luminosity increases with stellar mass (Fig.~\ref{massbat}). The lowest quartile stellar mass has $\log$ L$_{14-195\,\rm{keV}}= 43.07\pm0.88$ and the highest quartile stellar mass  has $\log$ L$_{14-195\,\rm{keV}}=43.72\pm0.36$.   In the lowest stellar mass quartile 34\% of sources have $\log$ L$_{14-195\,\rm{keV}}<$43 while in the highest stellar mass quartile none of the sources have $\log$ L$_{14-195\,\rm{keV}}<$43.   A K-S test has $<$0.2\% probability that the distributions of ultra hard X-ray luminosities are the same for the lowest quartile and highest quartile of stellar mass.  In addition, for the average stellar masses, we find a linear correlation between $\log$ L$_{14-195\,\rm{keV}}$ and $\log$ M$_*$ with a slope of 0.62$\pm$0.14 and a less than 2\% probability that the values are uncorrelated.
	
\subsection{Tests of Unification Model}	

	We also tested the Unified Model of Seyferts using the BAT sample.  In this model, it is assumed that all AGN are the same types of objects so host galaxy properties such as color, star formation, and morphology should be independent of the Seyfert type or level of obscuration toward the central engine. 
	
\begin{figure} 
\centering 
\plotone{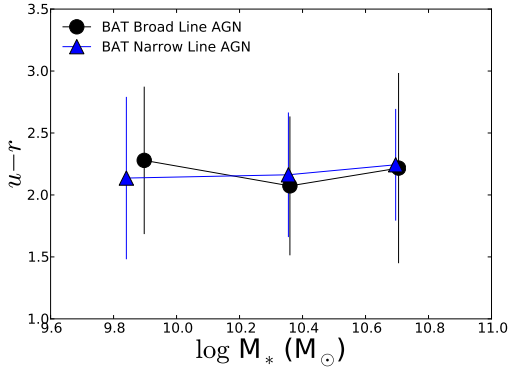} 
\caption{Comparison of average BAT AGN host galaxy colors separated by broad-line and narrow-line.  The error bars indicate standard deviations in each bin.   Petrosian $u-r$ by stellar mass (M$_{*}$) for broad-line and narrow-line AGN in the BAT sample.  Both narrow-line and broad-line AGN show similar host galaxy colors in agreement with AGN unification with P(K-S)= 99\% and 36\% for $u-r$ and $g-r$ respectively. } 
\label{agnunicolor}
\end{figure}

	We find that the host galaxy colors of narrow and broad-line AGN are the same in agreement with the unification model.  Both the $g-r$ and $u-r$ colors of broad-line AGN and NL AGN after GALFIT subtraction for AGN emission are very similar (Fig.~\ref{agnunicolor}).  For broad-line AGN, the mean $g-r$ is 0.66$\pm$0.15 and for NL, the mean $g-r$ is 0.68$\pm$0.12 with P(K-S)=43\% that the populations are the same.  In $u-r$, the color for broad-line AGN is 2.16$\pm$0.55 and NL AGN is 2.18$\pm$0.61 with P(K-S)=99\% that the populations are the same.
	
\begin{figure} 
\centering 
\plotone{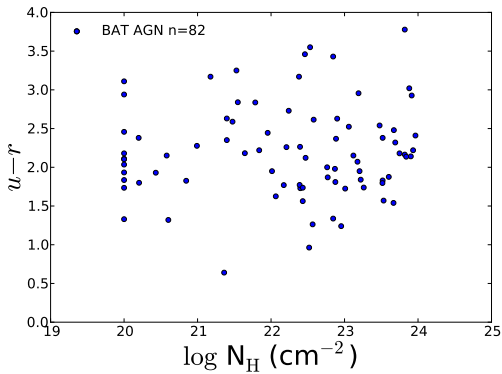} 
\caption{Scatter plot of X-ray column density vs.~host galaxy $u-r$. We see no correlation of X-ray column density with host galaxy color in agreement with AGN unification.  We also find no correlation in $g-r$.} 
\label{agnuninh}
\end{figure}

	An additional test of the Unified Model can be done by checking whether there is any correlation between color and column density in the BAT sample.  Column densities were obtained from the literature \citep{Winter:2008p7486,Winter:2009p79,Bassani:1999p5383,Noguchi:2010p7451} and the Tartarus database.    We see no correlation between column density and host galaxy color (Fig.~\ref{agnuninh}).  Since host galaxy color measures the relative amount of star formation, this suggest that there is no relation between X-ray column density (N$_{\mathrm{H}}$) and star formation. 

	We also investigated whether the NL and broad-line AGN have different rates of star formation in the far-IR.  We define a proxy for specific star formation rate as the logarithm of the ratio of 90 $\mu$m emission from \textit{AKARI} to stellar mass.  The mean value for this parameter for the narrow-line AGN is 33.6$\pm$0.4 erg s$^{-1} M_{\sun}^{-1}$ and for broad-line AGN is 33.4$\pm$0.4  erg s$^{-1} M_{\sun}^{-1}$.  A K-S test indicates a 35\% probability that the rates of specific star formation for the narrow and broad-line AGN are the same in agreement with the Unified Model of AGN.	
	
\begin{figure} 
\centering 
\plotone{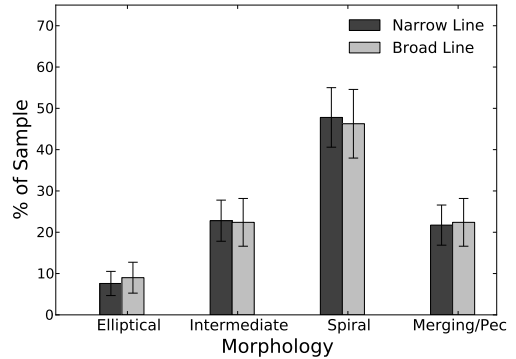}
\caption{Comparison of BAT AGN host galaxy morphologies separated by broad and narrow-line AGN.  The error bars represent 1$\sigma$ Poisson statistics.  We see no difference in the morphologies of the two samples in agreement with AGN unification.} 
\label{morphuni}
\end{figure}

	In addition, we compared the morphologies of broad and narrow-line AGN.  Because of the difficulty of determining morphology class for galaxies with very bright AGN, we limited our sample to galaxies with \%PS$_r<35$.   A plot of the percentage in each sample of different types of morphology can be found in Fig.~\ref{morphuni}.  We see no difference in the morphologies of broad and narrow-line AGN.
	
\begin{figure} 
\centering 
\plottwo{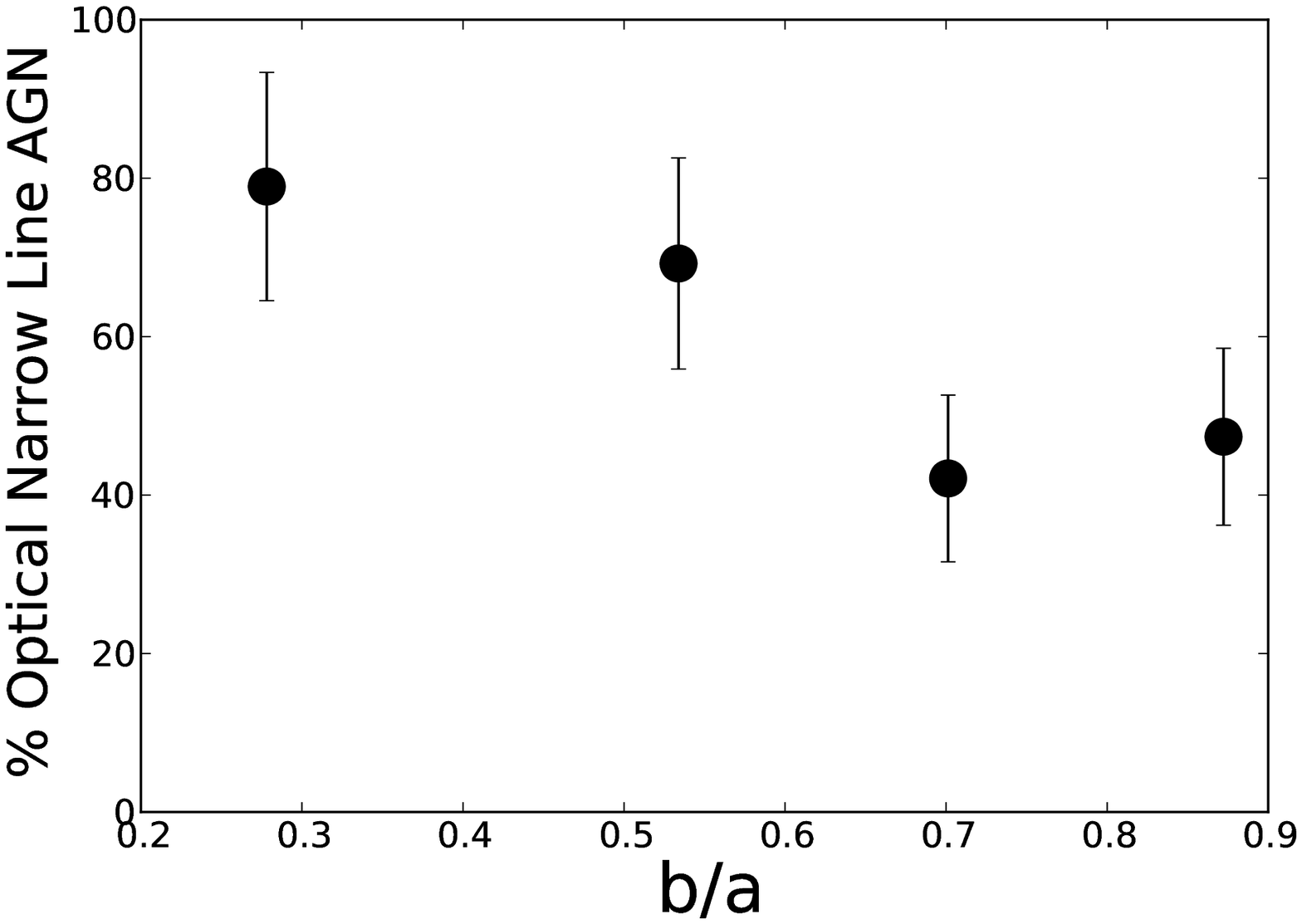}{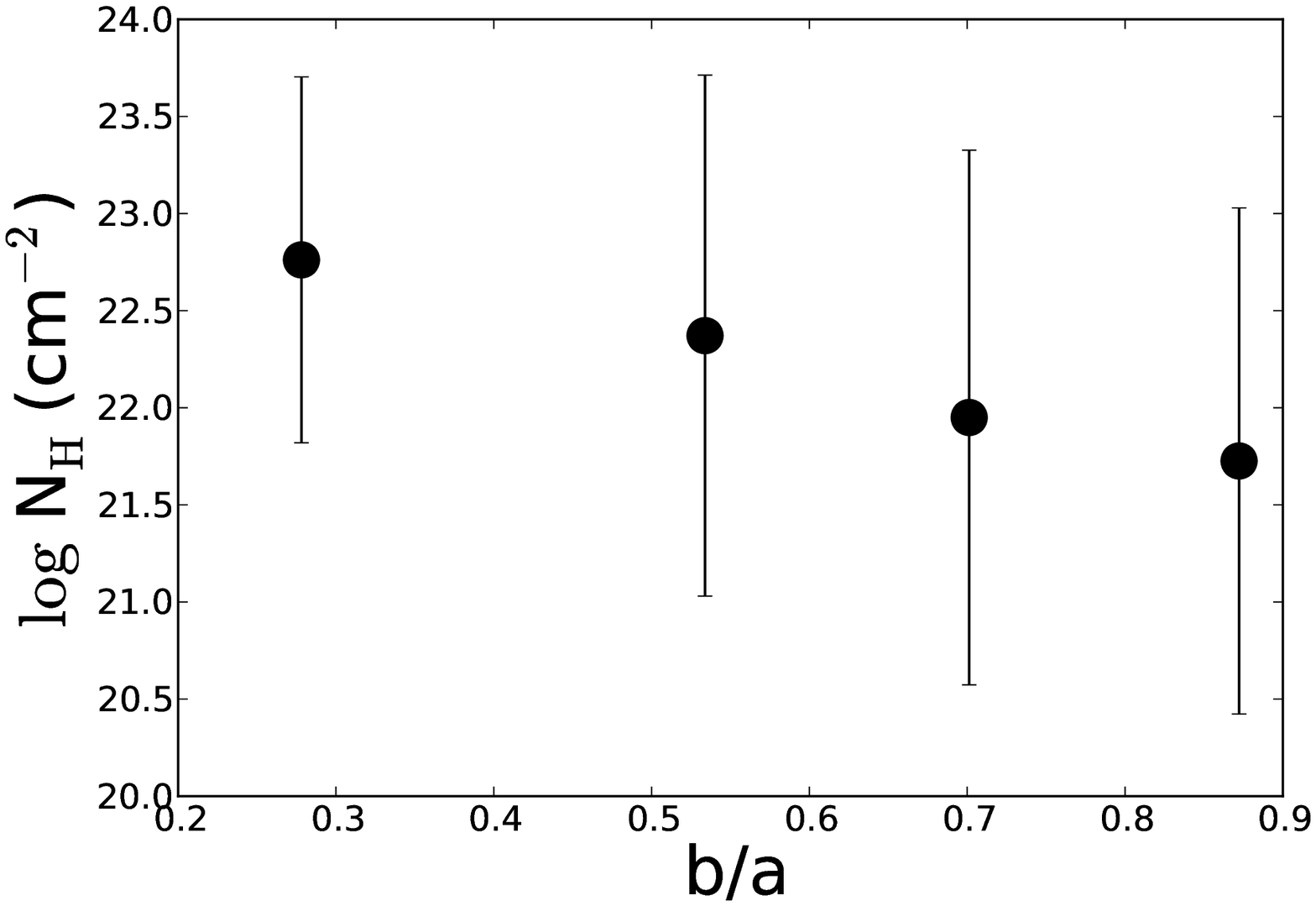}
\caption{{\em Left}:  Percent of narrow-line AGN vs. host galaxy inclination as measured by the axis ratio (b/a).  The error bars indicate standard deviations in each bin.  Galaxies with lower axis ratios tend to be more edge on.  We find more NL AGN in highly inclined systems with smaller axis ratios. {\em Right}:  X-ray column density vs.~host galaxy inclination as measured by the axis ratio (b/a).  The error bars indicate standard deviations in each bin.   Galaxies with higher inclinations have a higher mean X-ray column density.    A K-S test indicates a $<$5\% probability that the X-ray column densities from the lowest and highest distributions of axis ratios are from the same parent population. } 
\label{baxray}
\end{figure}

	Finally, we compared the axis ratios of BAT AGN with different optical classifications and obscuring column densities.  In the unified model of Seyferts, the observed X-ray spectra of Seyfert 2s are expected to have higher absorbing column density than Seyfert 1s due to an edge-on view of the obscuring torus.  However, edge-on spirals have been shown to have a geometrically thick layer of obscuring material in the host-galaxy planes that can also increase the absorbing column density \citep{Simcoe:1997p9312}.  We confirm this by finding more NL AGN in highly inclined systems with smaller axis ratios (b/a$<$0.4; Fig.~\ref{baxray}, left).  We also compared X-ray column density vs.~host galaxy inclination and found more inclined systems tend to have higher average X-ray column densities (Fig.~\ref{baxray}, right).  This finding confirms an earlier result from the smaller 9-month sample of BAT AGN \citep{Winter:2009p79}.
	
\section{Summary and Discussion}
	We have assembled the largest sample of ultra hard X-ray selected AGN with host galaxy optical data to date, with 185 AGN in total.  We have performed extensive modeling with GALFIT to effectively remove the AGN light from the optical images.  Using optical photometry, morphology, and spectroscopy, along with FIR emission we found:

\begin{enumerate}
\renewcommand{\theenumi}{(\roman{enumi})}
\renewcommand{\labelenumi}{\theenumi}
\item The BAT AGN galaxies are bluer in optical color than inactive galaxies or SDSS Seyferts of the same stellar mass.
\item We find a much higher incidence of spiral morphologies in BAT AGN compared to SDSS AGN or inactive galaxies.  Amongst massive galaxies ($\log$ M$_*>10.5$), the BAT AGN show a preference for spiral morphologies that is 5 to 10 times higher than SDSS AGN or inactive galaxies.  We also find that the bluer colors of BAT AGN can be accounted for by a higher fraction of mergers and spirals.
\item The BAT AGN have greatly enhanced 90 $\mu$m emission compared to inactive galaxies or SDSS Seyferts matched in redshift and stellar mass.
\item The BAT NL AGN have similar intrinsic [O \textsc{III}] $\lambda5007$ luminosities as NL SDSS Seyferts of the same redshift range.
\item The BAT AGN are found in the most massive host galaxies with high concentration indexes indicative of large bulge-to-disk ratios and large supermassive black holes.  
\item We also find that the average ultra hard X-ray luminosity increases with stellar mass and that BAT AGN in mergers have greater ultra hard X-ray emission than those in other morphological types.  This suggests a link between supermassive black hole growth and the mass of the host galaxy. 
\item In agreement with the Unified Model of AGN, we find the host galaxy colors and morphology are independent of X-ray column density and optical Seyfert classification.
\end{enumerate}
%on Galaxy Zoo and the RC3.  We also confirm the increased number of spirals based on the higher galaxy inclinations of BAT AGN.  Finally, we confirm the smaller number of ellipticals in the BAT AGN sample using SDSS photometry and spectroscopy catalog parameters

	These results indicate that host galaxy morphology is related to the activation and fueling of local AGN.  Ultra hard X-ray selected AGN are particularly associated with massive spiral galaxies and galaxy mergers.  These types of objects are generally associated with bluer colors, compared to the red massive early-type galaxies at similar stellar masses.  These observational results provide some evidence for an association between AGN activity and galaxy mergers (e.g., di Matteo et al.~2005), and also provide examples of AGN activity driven by the stochastic accretion of cold gas that should be more prominent among late-type systems \citep{Hopkins:2006p9574}.
	
	Recent simulations have also suggested a transition between the fueling mechanisms of AGN with nonmerger events predominantly powering lower luminosity AGN and merger-induced fueling dominant in more luminous quasars  \citep{Hopkins:2009p10071}.  We may be seeing evidence of this transition in our sample of BAT AGN that is powered both through merger events and less powerful nonmergers such as accretion of cold gas in late type systems.  In support of this, we find that BAT AGN in mergers have a greater ultra hard X-ray emission than those in other morphological types.  However, only a very small fraction (5/185) of BAT AGN in this sample are above the minimum bolometric luminosity associated with quasars (L$_{\mathrm{bol}}$$>$$10^{45}$ erg s$^{-1}$).  These results suggest that the process of merging may be important for powering more moderate luminosity AGN as well (see also Koss et al.~2010).
	
	In interpreting the results of an X-ray flux limited survey, it is useful to remember that the observed flux is a product of the black hole mass and accretion rate.  On average, more massive galaxies will tend to have higher mass black holes that will produce a larger average X-ray flux than smaller galaxies with on average smaller black holes.  However, among massive galaxies, elliptical morphologies are much more common than spirals, yet we find the most luminous hard X-ray AGN almost exclusively in spiral morphologies.  This suggests that spiral morphologies must have higher accretion rates than elliptical morphologies.  This finding is in agreement with recent theoretical predictions that suggest that only spirals typically have enough gas to trigger higher levels of radiatively efficient accretion in a geometrically thin disk \citep{Fanidakis:2011p10087}.  In order to understand this further, we are in the process of accurately measuring black hole masses to study the accretion rates for this sample.
	
	Previous optical surveys have found that AGN tend to be in massive galaxies \citep{Kauffmann:2003p2397}, occur along the red sequence \citep{Westoby:2007p8467}, and tend to have similar numbers of galaxy mergers as inactive galaxies \citep{Li:2006p5063}.  However, in an ultra hard X-ray survey of AGN, we find that AGN host galaxies are bluer than inactive galaxies with higher numbers of massive spirals and galaxy mergers.   We do not find observational evidence that the AGN suppresses star formation.  
	
	It is surprising that the optical morphologies and colors of ultra hard X-ray selected AGN are so different than emission line selected Type 2 Seyferts given their similar bolometric luminosity as measured in [O III].  However, these results are consistent with recent Spitzer surveys that have found that the AGN detection rate in late-type galaxies and mergers is much larger than what optical spectroscopic observations suggest \citep{Satyapal:2008p9525,Goulding:2009p6170, Veilleux:2009p7250}.  Finally, studies of X-ray selected AGN at higher redshifts, have also found a significant population of AGN classified as star forming using emission line diagnostics (Yan et al.~2011, accepted).

  	In the BAT AGN sample, there are several results that suggest optical emission line classification may be biased against late-type galaxies and mergers. In this study, we found that the axis ratios of BAT AGN are in general more inclined and have greater levels of internal extinction than comparable SDSS AGN.  This extinction could obscure or dilute the narrow-line region and cause AGN galaxies to be misclassified as star forming regions.  This finding is also in agreement with a previous analysis of BAT AGN that found optical emission line diagnostics preferentially misclassify merging AGN because of optical extinction and dilution by star formation \citep{Koss:2010p7366}. 
	
	Another possibility is that the BAT AGN may be much more intrinsically luminous than their [O III] emission suggests.  Since the majority of BAT AGN either have broad lines or are NL AGN that are correctly classified as Seyferts, yet are found to have much greater hard X-ray luminosities, this must be an important factor.  In support of this, two studies of BAT AGN have found a very weak correlation between the [O III] and hard X-ray luminosity and that BAT AGN have additional reddening of the narrow line region not accounted for in optical studies \citep{Winter:2010p6825,Melendez:2008p3807}.  This is also supported by the much greater number of narrow-line SDSS Seyferts compared to hard X-ray selected AGN.  In the SDSS survey area, there are 24 optical emission line selected narrow-line Seyferts detected for each hard X-ray AGN at the same redshift.  Some of these undetected sources may be heavily absorbed Compton-Thick AGN missed in the hard X-rays, but even the highest estimates expect only $\approx50\%$ of local narrow-line AGN are Compton~Thick \citep{Risaliti:1999p12126}. If the BAT AGN are intrinsically more luminous than emission line selected AGN, this may explain their higher rates of mergers and enhanced FIR emission.    We are currently in the process of assembling a larger survey of optical spectra of BAT AGN to better understand optical and X-ray measures of intrinsic luminosity.
		
\section{Acknowledgements}

We thank the anonymous referee for helpful comments that improved the presentation of this work.  We acknowledge the work that the $Swift$ BAT team has done to make this work possible. We are grateful to Meg Urry for useful discussion and suggestions. M. K. also acknowledges support through a Japanese Society for the Promotion of Science Fellowship, a Maryland Senatorial Scholarship, and a NASA graduate fellowship.   L.W. acknowledges support through a Hubble Fellowship from the Space Telescope Science Institute.  S.V. acknowledges support from a Senior Award from the Alexander von Humboldt Foundation and thanks the host institution, MPE Garching, where some of this work was performed.  The Kitt Peak National Observatory observations were obtained using MD-TAC time as part of the thesis of M. K. at the University of Maryland (programs 0417, 0393, and 0339).  Kitt Peak National Observatory, National Optical Astronomy Observatory, is operated by the Association of Universities for Research in Astronomy (AURA) under cooperative agreement with the National Science Foundation.  Based on observations from \textit{AKARI}, a JAXA project with the participation of ESA.  This research used the Tartarus database, created by Paul O'Neill and Kirpal Nandra at Imperial College London, and Jane Turner at NASA/GSFC. Tartarus is supported by funding from PPARC, and NASA grants NAG5-7385 and NAG5-7067.  Finally, this research used the NASA/IPAC Extragalactic Database (NED) which is operated by the Jet Propulsion Laboratory, California Institute of Technology, under contract with the NASA.

%and L.W. (programs 0322 and 0107)
{\it Facilities:}  \facility{Swift}, \facility{Sloan}, \facility{KPNO:2.1m}, \facility{IRAS}, \facility{AKARI}
\bibliographystyle{/Applications/astronat/apj/apj}
\bibliography{/Applications/astronat/bibfinal}

\begin{thebibliography}{71}
\expandafter\ifx\csname natexlab\endcsname\relax\def\natexlab#1{#1}\fi

\bibitem[{Ajello {et~al.}(2008)Ajello, Rau, Greiner, Kanbach, Salvato, Strong,
  Barthelmy, Gehrels, Markwardt, \& Tueller}]{Ajello:2008p9070}
Ajello, M., {et~al.} 2008, ApJ, 673, 96

\bibitem[{Bassani {et~al.}(1999)Bassani, Dadina, Maiolino, Salvati, Risaliti,
  della Ceca, Matt, \& Zamorani}]{Bassani:1999p5383}
Bassani, L., Dadina, M., Maiolino, R., Salvati, M., Risaliti, G., della Ceca,
  R., Matt, G., \& Zamorani, G. 1999, ApJSS, 121, 473

\bibitem[{Bernardi {et~al.}(2003)Bernardi, Sheth, Annis, Burles, Eisenstein,
  Finkbeiner, Hogg, Lupton, Schlegel, SubbaRao, Bahcall, Blakeslee, Brinkmann,
  Castander, Connolly, Csabai, Doi, Fukugita, Frieman, Heckman, Hennessy,
  Ivezi{\'c}, Knapp, Lamb, McKay, Munn, Nichol, Okamura, Schneider, Thakar, \&
  York}]{Bernardi:2003p9326}
Bernardi, M., {et~al.} 2003, AJ, 125, 1849

\bibitem[{Bertin \& Arnouts(1996)}]{Bertin:1996p9137}
Bertin, E., \& Arnouts, S. 1996, A\&AS, 117, 393

\bibitem[{Blanton {et~al.}(2001)Blanton, Dalcanton, Eisenstein, Loveday,
  Strauss, SubbaRao, Weinberg, Anderson, Annis, Bahcall, Bernardi, Brinkmann,
  Brunner, Burles, Carey, Castander, Connolly, Csabai, Doi, Finkbeiner,
  Friedman, Frieman, Fukugita, Gunn, Hennessy, Hindsley, Hogg, Ichikawa,
  Ivezi{\'c}, Kent, Knapp, Lamb, Leger, Long, Lupton, McKay, Meiksin, Merelli,
  Munn, Narayanan, Newcomb, Nichol, Okamura, Owen, Pier, Pope, Postman, Quinn,
  Rockosi, Schlegel, Schneider, Shimasaku, Siegmund, Smee, Snir, Stoughton,
  Stubbs, Szalay, Szokoly, Thakar, Tremonti, Tucker, Uomoto, Berk, Vogeley,
  Waddell, Yanny, Yasuda, \& York}]{Blanton:2001p8687}
Blanton, M.~R., {et~al.} 2001, AJ, 121, 2358

\bibitem[{Blanton \& Roweis(2007)}]{Blanton:2007p3139}
Blanton, M.~R., \& Roweis, S. 2007, AJ, 133, 734

\bibitem[{Bruzual \& Charlot(2003)}]{Bruzual:2003p1640}
Bruzual, G., \& Charlot, S. 2003, MNRAS, 344, 1000

\bibitem[{Calzetti {et~al.}(2000)Calzetti, Armus, Bohlin, Kinney, Koornneef, \&
  Storchi-Bergmann}]{Calzetti:2000p9589}
Calzetti, D., Armus, L., Bohlin, R., Kinney, A., Koornneef, J., \&
  Storchi-Bergmann, T. 2000, ApJ, 533, 682

\bibitem[{Cardelli {et~al.}(1989)Cardelli, Clayton, \&
  Mathis}]{Cardelli:1989p1821}
Cardelli, J.~A., Clayton, G.~C., \& Mathis, J.~S. 1989, ApJ, 345, 245

\bibitem[{Chabrier(2003)}]{Chabrier:2003p9594}
Chabrier, G. 2003, PASJ, 115, 763

\bibitem[{Dekel {et~al.}(2009)Dekel, Birnboim, Engel, Freundlich, Goerdt,
  Mumcuoglu, Neistein, Pichon, Teyssier, \& Zinger}]{Dekel:2009p9116}
Dekel, A., {et~al.} 2009, Nature, 457, 451

\bibitem[{di~Matteo {et~al.}(2005)di~Matteo, Springel, \&
  Hernquist}]{DiMatteo:2005p5934}
di~Matteo, T., Springel, V., \& Hernquist, L. 2005, Nature, 433, 604

\bibitem[{Fanidakis {et~al.}(2011)Fanidakis, Baugh, Benson, Bower, Cole, Done,
  \& Frenk}]{Fanidakis:2011p10087}
Fanidakis, N., Baugh, C.~M., Benson, A.~J., Bower, R.~G., Cole, S., Done, C.,
  \& Frenk, C.~S. 2011, MNRAS, 410, 53

\bibitem[{Gavazzi {et~al.}(2000)Gavazzi, Franzetti, Scodeggio, Boselli, \&
  Pierini}]{Gavazzi:2000p9275}
Gavazzi, G., Franzetti, P., Scodeggio, M., Boselli, A., \& Pierini, D. 2000,
  A\&A, 361, 863

\bibitem[{Goulding \& Alexander(2009)}]{Goulding:2009p6170}
Goulding, A.~D., \& Alexander, D.~M. 2009, MNRAS, 398, 1165

\bibitem[{Haubler {et~al.}(2008)Haubler, Barden, \&
  McIntosh}]{Haubler:2008p734}
Haubler, B., Barden, M., \& McIntosh, D.~H. 2008, AIPC, 1082, 137

\bibitem[{Heckman(1980)}]{Heckman:1980p9959}
Heckman, T.~M. 1980, A\&A, 87, 152

\bibitem[{Hickox {et~al.}(2009)Hickox, Jones, Forman, Murray, Kochanek,
  Eisenstein, Jannuzi, Dey, Brown, Stern, Eisenhardt, Gorjian, Brodwin,
  Narayan, Cool, Kenter, Caldwell, \& Anderson}]{Hickox:2009p707}
Hickox, R.~C., {et~al.} 2009, eprint arXiv, 0901, 4121

\bibitem[{Ho {et~al.}(1997)Ho, Filippenko, Sargent, \& Peng}]{Ho:1997p5224}
Ho, L.~C., Filippenko, A.~V., Sargent, W. L.~W., \& Peng, C.~Y. 1997, ApJS,
  112, 391

\bibitem[{Hopkins \& Hernquist(2006)}]{Hopkins:2006p9574}
Hopkins, P.~F., \& Hernquist, L. 2006, ApJSS, 166, 1

\bibitem[{Hopkins \& Hernquist(2009)}]{Hopkins:2009p10071}
---. 2009, ApJ, 694, 599

\bibitem[{Kauffmann {et~al.}(2003)Kauffmann, Heckman, Tremonti, Brinchmann,
  Charlot, White, Ridgway, Brinkmann, Fukugita, Hall, Ivezi{\'c}, Richards, \&
  Schneider}]{Kauffmann:2003p2397}
Kauffmann, G., {et~al.} 2003, MNRAS, 346, 1055

\bibitem[{Kewley {et~al.}(2006)Kewley, Groves, Kauffmann, \&
  Heckman}]{Kewley:2006p1554}
Kewley, L.~J., Groves, B., Kauffmann, G., \& Heckman, T. 2006, MNRAS, 372, 961

\bibitem[{Kewley {et~al.}(2001)Kewley, Heisler, Dopita, \&
  Lumsden}]{Kewley:2001p5316}
Kewley, L.~J., Heisler, C.~A., Dopita, M.~A., \& Lumsden, S. 2001, ApJSS, 132,
  37

\bibitem[{Kim {et~al.}(2008)Kim, Ho, Peng, Barth, Im, Martini, \&
  Nelson}]{Kim:2008p1462}
Kim, M., Ho, L.~C., Peng, C.~Y., Barth, A.~J., Im, M., Martini, P., \& Nelson,
  C.~H. 2008, ApJ, 687, 767

\bibitem[{Kirhakos \& Steiner(1990)}]{Kirhakos:1990p9311}
Kirhakos, S.~D., \& Steiner, J.~E. 1990, AJ, 99, 1435

\bibitem[{Koss {et~al.}(2011)Koss, Mushotzky, Treister, Veilleux, Vasudevan,
  Miller, Sanders, Schawinski, \& Trippe}]{Koss:2011p12483}
Koss, M., {et~al.} 2011, ApJL, 735, L42

\bibitem[{Koss {et~al.}(2010)Koss, Mushotzky, Veilleux, \&
  Winter}]{Koss:2010p7366}
Koss, M., Mushotzky, R., Veilleux, S., \& Winter, L. 2010, ApJL, 716, L125

\bibitem[{Kriek {et~al.}(2009)Kriek, van Dokkum, Labb{\'e}, Franx, Illingworth,
  Marchesini, \& Quadri}]{Kriek:2009p9587}
Kriek, M., van Dokkum, P.~G., Labb{\'e}, I., Franx, M., Illingworth, G.~D.,
  Marchesini, D., \& Quadri, R.~F. 2009, ApJ, 700, 221

\bibitem[{Levine {et~al.}(1984)Levine, Lang, Lewin, Primini, Dobson, Doty,
  Hoffman, Howe, Scheepmaker, Wheaton, Matteson, Baity, Gruber, Knight, Nolan,
  Pelling, Rothschild, \& Peterson}]{Levine:1984p9096}
Levine, A.~M., {et~al.} 1984, ApJSS, 54, 581

\bibitem[{Li {et~al.}(2006)Li, Kauffmann, Wang, White, Heckman, \&
  Jing}]{Li:2006p5063}
Li, C., Kauffmann, G., Wang, L., White, S. D.~M., Heckman, T.~M., \& Jing,
  Y.~P. 2006, MNRAS, 373, 457

\bibitem[{Lintott {et~al.}(2008)Lintott, Schawinski, Slosar, Land, Bamford,
  Thomas, Raddick, Nichol, Szalay, Andreescu, Murray, \&
  Vandenberg}]{Lintott:2008p9199}
Lintott, C.~J., {et~al.} 2008, MNRAS, 389, 1179

\bibitem[{Lupton {et~al.}(2004)Lupton, Blanton, Fekete, Hogg, O'Mullane,
  Szalay, \& Wherry}]{Lupton:2004p7328}
Lupton, R., Blanton, M.~R., Fekete, G., Hogg, D.~W., O'Mullane, W., Szalay, A.,
  \& Wherry, N. 2004, PASJ, 116, 133

\bibitem[{Magorrian {et~al.}(1998)Magorrian, Tremaine, Richstone, Bender,
  Bower, Dressler, Faber, Gebhardt, Green, Grillmair, Kormendy, \&
  Lauer}]{Magorrian:1998p9015}
Magorrian, J., {et~al.} 1998, AJ, 115, 2285

\bibitem[{Malizia {et~al.}(2009)Malizia, Stephen, Bassani, Bird, Panessa, \&
  Ubertini}]{Malizia:2009p8784}
Malizia, A., Stephen, J.~B., Bassani, L., Bird, A.~J., Panessa, F., \&
  Ubertini, P. 2009, MNRAS, 399, 944

\bibitem[{Maller {et~al.}(2009)Maller, Berlind, Blanton, \&
  Hogg}]{Maller:2009p9276}
Maller, A.~H., Berlind, A.~A., Blanton, M.~R., \& Hogg, D.~W. 2009, ApJ, 691,
  394

\bibitem[{Masters {et~al.}(2010)Masters, Nichol, Bamford, Mosleh, Lintott,
  Andreescu, Edmondson, Keel, Murray, Raddick, Schawinski, Slosar, Szalay,
  Thomas, \& Vandenberg}]{Masters:2010p8988}
Masters, K.~L., {et~al.} 2010, MNRAS, 404, 792

\bibitem[{Mel{\'e}ndez {et~al.}(2008)Mel{\'e}ndez, Kraemer, Armentrout, Deo,
  Crenshaw, Schmitt, Mushotzky, Tueller, Markwardt, \&
  Winter}]{Melendez:2008p3807}
Mel{\'e}ndez, M., {et~al.} 2008, ApJ, 682, 94

\bibitem[{Mink(1996)}]{Mink:1996p9129}
Mink, D.~J. 1996, Astronomical Data Analysis Software and Systems V, 101, 96

\bibitem[{Mulchaey \& Regan(1997)}]{Mulchaey:1997p9021}
Mulchaey, J.~S., \& Regan, M.~W. 1997, ApJL, 482, L135

\bibitem[{Mullaney {et~al.}(2011)Mullaney, Alexander, Goulding, \&
  Hickox}]{Mullaney:2011p12174}
Mullaney, J.~R., Alexander, D.~M., Goulding, A.~D., \& Hickox, R.~C. 2011,
  MNRAS, 414, 1082

\bibitem[{Mushotzky(2004)}]{Mushotzky:2004p8576}
Mushotzky, R. 2004, Supermassive Black Holes in the Distant Universe. Edited by
  Amy J. Barger, 308, 53

\bibitem[{Netzer {et~al.}(2007)Netzer, Lutz, Schweitzer, Contursi, Sturm,
  Tacconi, Veilleux, Kim, Rupke, Baker, Dasyra, Mazzarella, \&
  Lord}]{Netzer:2007p1714}
Netzer, H., {et~al.} 2007, ApJ, 666, 806

\bibitem[{Noguchi {et~al.}(2010)Noguchi, Terashima, Ishino, Hashimoto, Koss,
  Ueda, \& Awaki}]{Noguchi:2010p7451}
Noguchi, K., Terashima, Y., Ishino, Y., Hashimoto, Y., Koss, M., Ueda, Y., \&
  Awaki, H. 2010, ApJ, 711, 144

\bibitem[{Patton \& Atfield(2008)}]{Patton:2008p5024}
Patton, D.~R., \& Atfield, J.~E. 2008, ApJ, 685, 235

\bibitem[{Peng {et~al.}(2002)Peng, Ho, Impey, \& Rix}]{Peng:2002p5550}
Peng, C.~Y., Ho, L.~C., Impey, C.~D., \& Rix, H.-W. 2002, AJ, 124, 266

\bibitem[{Pierce {et~al.}(2010)Pierce, Lotz, Primack, Rosario, Griffith,
  Conselice, Faber, Koo, Coil, Salim, Koekemoer, Laird, Ivison, \&
  Yan}]{Pierce:2010p8180}
Pierce, C.~M., {et~al.} 2010, MNRAS, 405, 718

\bibitem[{Richards {et~al.}(2006)Richards, Lacy, Storrie-Lombardi, Hall,
  Gallagher, Hines, Fan, Papovich, Berk, Trammell, Schneider, Vestergaard,
  York, Jester, Anderson, Budav{\'a}ri, \& Szalay}]{Richards:2006p3527}
Richards, G.~T., {et~al.} 2006, ApJSS, 166, 470

\bibitem[{Risaliti {et~al.}(1999)Risaliti, Maiolino, \&
  Salvati}]{Risaliti:1999p12126}
Risaliti, G., Maiolino, R., \& Salvati, M. 1999, ApJ, 522, 157

\bibitem[{Satyapal {et~al.}(2008)Satyapal, Vega, Dudik, Abel, \&
  Heckman}]{Satyapal:2008p9525}
Satyapal, S., Vega, D., Dudik, R.~P., Abel, N.~P., \& Heckman, T. 2008, ApJ,
  677, 926

\bibitem[{Schawinski {et~al.}(2010)Schawinski, Urry, Virani, Coppi, Bamford,
  Treister, Lintott, Sarzi, Keel, Kaviraj, Cardamone, Masters, Ross, Andreescu,
  Murray, Nichol, Raddick, Slosar, Szalay, Thomas, \&
  Vandenberg}]{Schawinski:2010p6049}
Schawinski, K., {et~al.} 2010, ApJ, 711, 284

\bibitem[{Schawinski {et~al.}(2009)Schawinski, Virani, Simmons, Urry, Treister,
  Kaviraj, \& Kushkuley}]{Schawinski:2009p1181}
Schawinski, K., Virani, S., Simmons, B., Urry, C.~M., Treister, E., Kaviraj,
  S., \& Kushkuley, B. 2009, ApJL, 692, L19

\bibitem[{Schlegel {et~al.}(1998)Schlegel, Finkbeiner, \&
  Davis}]{Schlegel:1998p9121}
Schlegel, D.~J., Finkbeiner, D.~P., \& Davis, M. 1998, ApJ, 500, 525

\bibitem[{Silverman {et~al.}(2008)Silverman, Mainieri, Lehmer, Alexander,
  Bauer, Bergeron, Brandt, Gilli, Hasinger, Schneider, Tozzi, Vignali,
  Koekemoer, Miyaji, Popesso, Rosati, \& Szokoly}]{Silverman:2008p8462}
Silverman, J.~D., {et~al.} 2008, ApJ, 675, 1025

\bibitem[{Simcoe {et~al.}(1997)Simcoe, McLeod, Schachter, \&
  Elvis}]{Simcoe:1997p9312}
Simcoe, R., McLeod, K.~K., Schachter, J., \& Elvis, M. 1997, ApJ, 489, 615

\bibitem[{Simmons \& Urry(2008)}]{Simmons:2008p2188}
Simmons, B., \& Urry, C. 2008, ApJ, 683, 644

\bibitem[{Smith {et~al.}(2002)Smith, Tucker, Kent, Richmond, Fukugita,
  Ichikawa, ichi Ichikawa, Jorgensen, Uomoto, Gunn, Hamabe, Watanabe, Tolea,
  Henden, Annis, Pier, McKay, Brinkmann, Chen, Holtzman, Shimasaku, \&
  York}]{Smith:2002p9117}
Smith, J.~A., {et~al.} 2002, AJ, 123, 2121

\bibitem[{Stern {et~al.}(2005)Stern, Eisenhardt, Gorjian, Kochanek, Caldwell,
  Eisenstein, Brodwin, Brown, Cool, Dey, Green, Jannuzi, Murray, Pahre, \&
  Willner}]{Stern:2005p8982}
Stern, D., {et~al.} 2005, ApJ, 631, 163

\bibitem[{Tueller {et~al.}(2010)Tueller, Baumgartner, Markwardt, Skinner,
  Mushotzky, Ajello, Barthelmy, Beardmore, Brandt, Burrows, Chincarini,
  Campana, Cummings, Cusumano, Evans, Fenimore, Gehrels, Godet, Grupe, Holland,
  Kennea, Krimm, Koss, Moretti, Mukai, Osborne, Okajima, Pagani, Page, Palmer,
  Parsons, Schneider, Sakamoto, Sambruna, Sato, Stamatikos, Stroh, Ukwata, \&
  Winter}]{Tueller:2010p6018}
Tueller, J., {et~al.} 2010, ApJS, 186, 378

\bibitem[{Tueller {et~al.}(2008)Tueller, Mushotzky, Barthelmy, Cannizzo,
  Gehrels, Markwardt, Skinner, \& Winter}]{Tueller:2008p2710}
Tueller, J., Mushotzky, R.~F., Barthelmy, S., Cannizzo, J.~K., Gehrels, N.,
  Markwardt, C.~B., Skinner, G.~K., \& Winter, L.~M. 2008, ApJ, 681, 113

\bibitem[{Vasudevan {et~al.}(2009)Vasudevan, Mushotzky, Winter, \&
  Fabian}]{Vasudevan:2009p7223}
Vasudevan, R.~V., Mushotzky, R.~F., Winter, L.~M., \& Fabian, A.~C. 2009,
  MNRAS, 399, 1553

\bibitem[{Veilleux {et~al.}(2006)Veilleux, Kim, Peng, Ho, Tacconi, Dasyra,
  Genzel, Lutz, \& Sanders}]{Veilleux:2006p1128}
Veilleux, S., {et~al.} 2006, ApJ, 643, 707

\bibitem[{Veilleux {et~al.}(2009{\natexlab{a}})Veilleux, Kim, Rupke, Peng,
  Tacconi, Genzel, Lutz, Sturm, Contursi, Schweitzer, Dasyra, Ho, Sanders, \&
  Burkert}]{Veilleux:2009p9544}
---. 2009{\natexlab{a}}, ApJ, 701, 587

\bibitem[{Veilleux {et~al.}(2009{\natexlab{b}})Veilleux, Rupke, Kim, Genzel,
  Sturm, Lutz, Contursi, Schweitzer, Tacconi, Netzer, Sternberg, Mihos, Baker,
  Mazzarella, Lord, Sanders, Stockton, Joseph, \& Barnes}]{Veilleux:2009p7250}
---. 2009{\natexlab{b}}, ApJS, 182, 628

\bibitem[{Weedman(1977)}]{Weedman:1977p8832}
Weedman, D.~W. 1977, ARA\&A, 15, 69

\bibitem[{Westoby {et~al.}(2007)Westoby, Mundell, \&
  Baldry}]{Westoby:2007p8467}
Westoby, P.~B., Mundell, C.~G., \& Baldry, I.~K. 2007, MNRAS, 382, 1541

\bibitem[{Winter {et~al.}(2008)Winter, Mushotzky, Tueller, \&
  Markwardt}]{Winter:2008p7486}
Winter, L., Mushotzky, R., Tueller, J., \& Markwardt, C. 2008, ApJ, 674, 686

\bibitem[{Winter {et~al.}(2010)Winter, Lewis, Koss, Veilleux, Keeney, \&
  Mushotzky}]{Winter:2010p6825}
Winter, L.~M., Lewis, K.~T., Koss, M., Veilleux, S., Keeney, B., \& Mushotzky,
  R.~F. 2010, ApJ, 710, 503

\bibitem[{Winter {et~al.}(2009)Winter, Mushotzky, Reynolds, \&
  Tueller}]{Winter:2009p79}
Winter, L.~M., Mushotzky, R.~F., Reynolds, C.~S., \& Tueller, J. 2009, ApJ,
  690, 1322

\end{thebibliography}
\expandafter\ifx\csname natexlab\endcsname\relax\def\natexlab#1{#1}\fi

\appendix
\section{AGN Subtraction and GALFIT Analysis}

 	The AGN color is bluer than the host galaxy, so it is important to accurately subtract the AGN light before doing photometry of the host galaxy.  Two-dimensional surface brightness fitting was done using GALFIT \citep{Peng:2002p5550} in the  \textit{ugriz} band to measure and subtract the AGN light.  The program can simultaneously fit an arbitrary number of components using $\chi^{2}$ minimization to determine the best-fit parameters.  Our choice of GALFIT is based on the recent comparison of GIM2D vs.~GALFIT which showed better fitting results and stability in finding solutions \citep{Haubler:2008p734}.  While the median atmospheric seeing of our sample was only $\approx$1.5\arcsec, since the sample is at a very low redshift, this ground-based optical imaging is comparable or even superior to the best Hubble Space Telescope (HST) images at high-redshift (z$>$0.5). 	
	
	Initial estimates of galaxy magnitude, position, position angle, axis ratio, and half-light radius were provided using SExtractor following the GALAPOGOS routine \citep{Haubler:2008p734}.  A point source (PS) was used to fit the AGN light.  The Point Source Function (PSF) was modeled using five coadded bright stars in the same image field as the galaxy.   An initial run of GALFIT was done using only a PS component to replace the SExtractor inputs for central position and PS magnitudes. Sky background estimates were made using SDSS sky values or from SExtractor. 
	
	To model the galaxy light we used the S\'{e}rsic profile, which is an exponential function with a variable half-light radius and an exponential parameter, n, called the S\'{e}rsic index (S\'{e}rsic 1968).  For n$=$1, the S\'{e}rsic profile is the same as an exponential disk model.  When n$=$4, the S\'{e}rsic profile is the same as a de Vaucouleurs bulge.  Other authors have used different fixed and floating S\'{e}rsic index models including a fixed bulge (n=4), a fixed disk and bulge (n=1 and n=4), and a floating S\'{e}rsic index \citep{Veilleux:2006p1128,Veilleux:2009p9544,Schawinski:2009p1181}. While a detailed study of the most effective way to measure the AGN and galaxy light has been done for simulated HST images \citep{Simmons:2008p2188, Kim:2008p1462,Pierce:2010p8180}, little has been done for ground-based images similar to the current study.  Therefore, to determine the best modeling approach with GALFIT and the associated error, we simulated AGN galaxies for both our Kitt Peak and SDSS images. 
	
	To determine the best model to measure the AGN and galaxy light, we simulated broad-line AGN galaxies by adding bright stars to the centers of images of inactive galaxies and NL AGN.  We randomly selected one star from our images to use as the simulated AGN PS and placed it in the center of the galaxy.  Since the SDSS and Kitt Peak data had different PSFs and exposure times, we tested them separately.  To test the SDSS data, we selected 15 inactive galaxies from the SDSS catalog which matched in redshift, color, and apparent magnitude to our Seyfert 2 galaxies with $0.025<z<0.05$.  These galaxies are the most distant in our sample and have the poorest resolution, so PS subtraction leads to large errors; they therefore serve as a worst case scenario for our sample.  For the Kitt Peak images we chose a sample of 10 of the BAT Seyfert 2 galaxies with the same redshift range.   For each of the simulated AGN galaxies we added the star at incremental percentage (\%PS$_{r}$)  values of total (AGN and galaxy) light in the $r$ band.    In total, to test them in each filter, we created 300 simulated AGN galaxies and ran GALFIT 3500 times.   We then used these simulated AGN galaxies to test the effectiveness of GALFIT with different models.    
	
\begin{figure} 
\centering 
\plotone{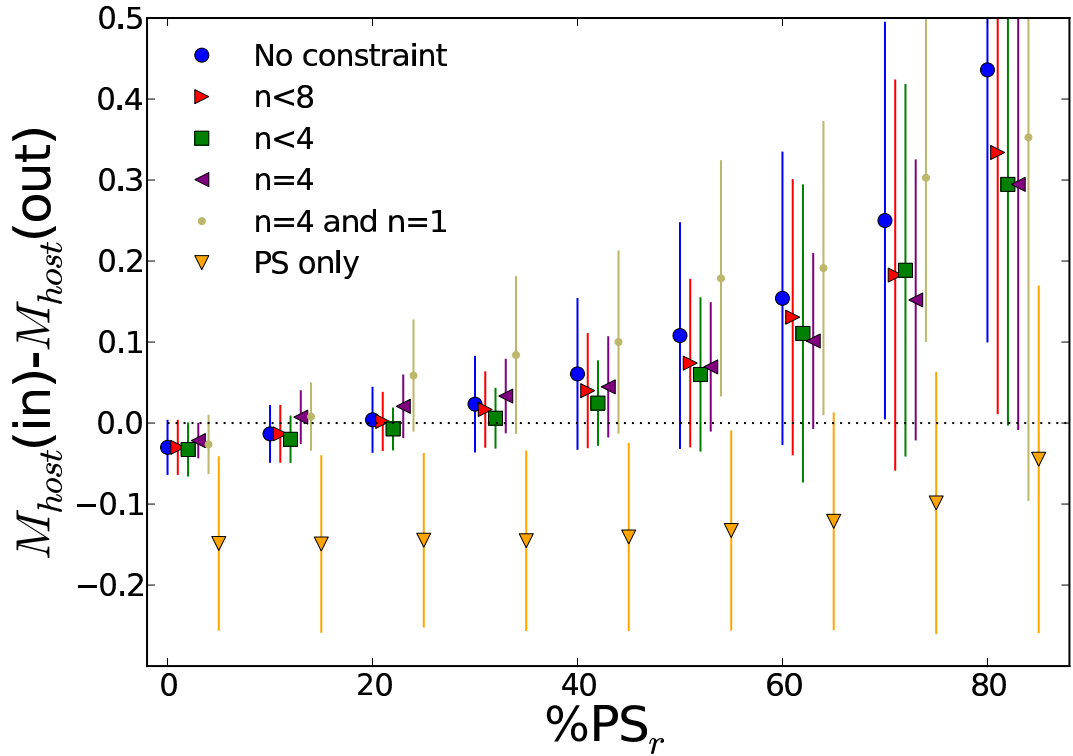} 
\caption{To determine the best models to use for subtraction of the AGN light, we tested simulated AGN galaxies using different S\'{e}rsic models with a point source (PS) to fit the AGN light.   We varied the AGN to total light (\%PS$_{r}$) in the $r$ band to test how the models performed at increasing levels of AGN light.  A positive value indicates that the mean host galaxy light is overestimated.  The error bars indicate standard deviations in each bin.   As the (\%PS$_{r}$) increases the host galaxy light is overestimated by all of the S\'{e}rsic and PS models and the standard deviation increases.  We find that the model with a PS and floating S\'{e}rsic Index with n$<$4 or n$=4$ is most effective at measuring the AGN light based on a lower measured AGN contamination and smaller standard deviation.   }
\label{f15}
\end{figure}

	Fig.~\ref{f15} shows the simulation results for the different S\'{e}rsic models for increasing \%PS$_{r}$ light in the $r$ band.  We did not find a significant difference in GALFIT modeling using the Kitt Peak or SDSS samples, so these results include both samples.  We found an inaccurately modeled PSF will force GALFIT to converge to artificially high S\'{e}rsic indexes for the galaxy model.  This has also been found in simulated HST images \citep{Simmons:2008p2188, Kim:2008p1462}.   The PSF mismatch causes light from the host galaxy component to be artificially increased by effectively taking light from the AGN component.  This happens by inflating the galaxy S\'{e}rsic index.   This effect increases as the \%PS$_{r}$ increases. In addition, as we move towards larger \%PS$_{r}$, the associated standard deviation of error of the modeled galaxy light increases.  
	
	To accurately remove the AGN light, it is important to choose the best model.  We did this by finding the difference between the modeled galaxy light and the actual galaxy light.   When no PS component is used in the models, GALFIT still finds a faint AGN, so after subtraction, the modeled galaxy light component is fainter than the actual one.  When the host galaxy is brighter than the AGN in the $r$ band (\%PS$_{r}$$<$50), the worst model is to simply subtract a PS component to estimate the galaxy light because this overestimates the AGN light.  For the S\'{e}rsic and PS fitting models, the best model is to use an n$=$4 fixed bulge component and PS or n$<$4 variable S\'{e}rsic Index and PS since these models have the smallest average error and standard deviation of error.  The worst model is fitting with a disk(n=1) and bulge(n=4) or fitting with a variable S\'{e}rsic Index since these models have the largest average overestimation of galaxy light and have the largest standard deviation of error.  Simmons \& Urry (2008) also found this result using simulated HST images.  To avoid any systematic biases against disk-like systems we used the n$<$4 variable S\'{e}rsic Index although the fixed bulge (n$=$4) S\'{e}rsic model performed similarly.  An initial guess of 2.5 was used for the S\'{e}rsic Index.  The index was allowed to float in the model along with all other values other than the sky background.    

\begin{figure} 
\centering 
\includegraphics[width=12cm]{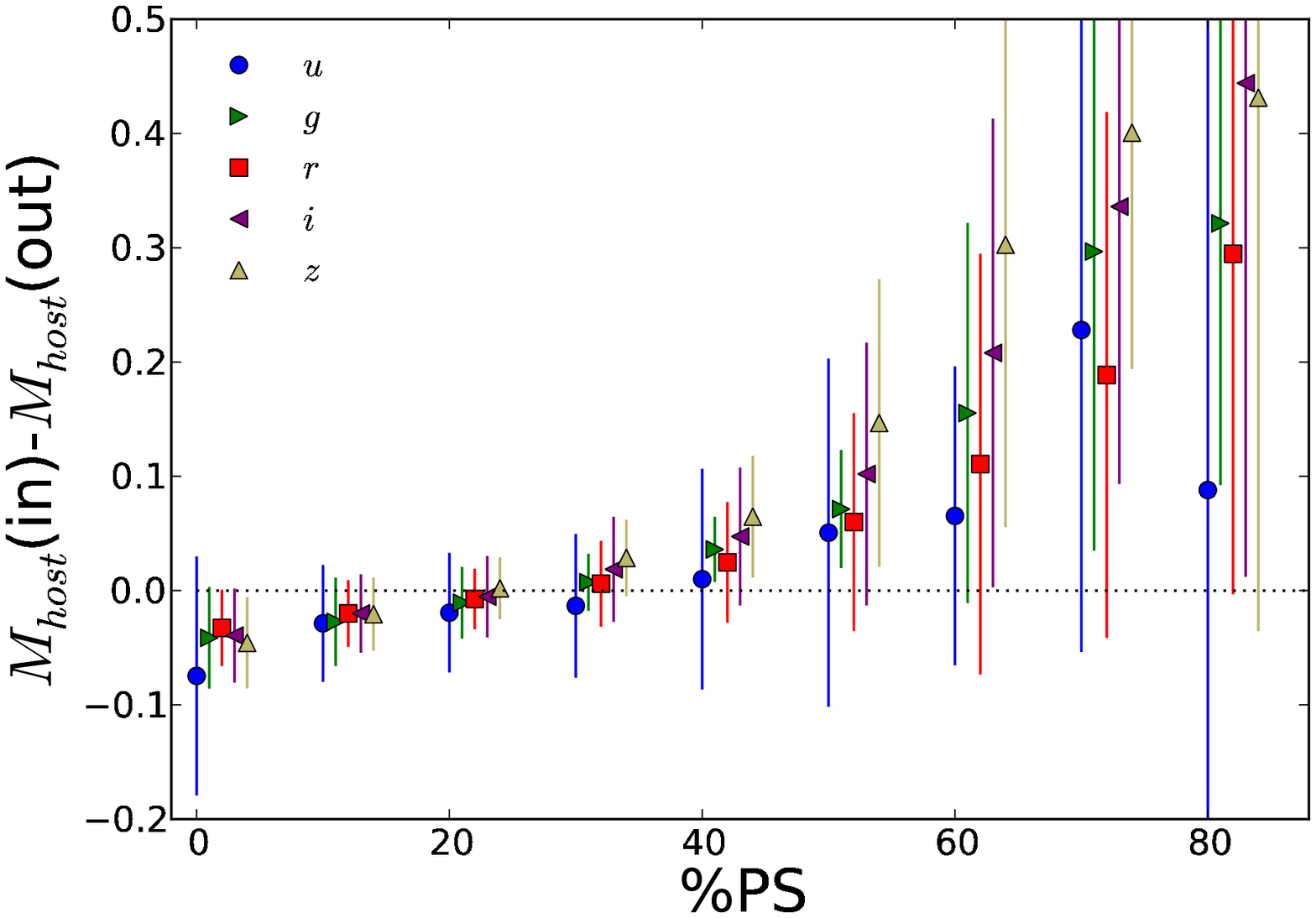} 
\includegraphics[width=12cm]{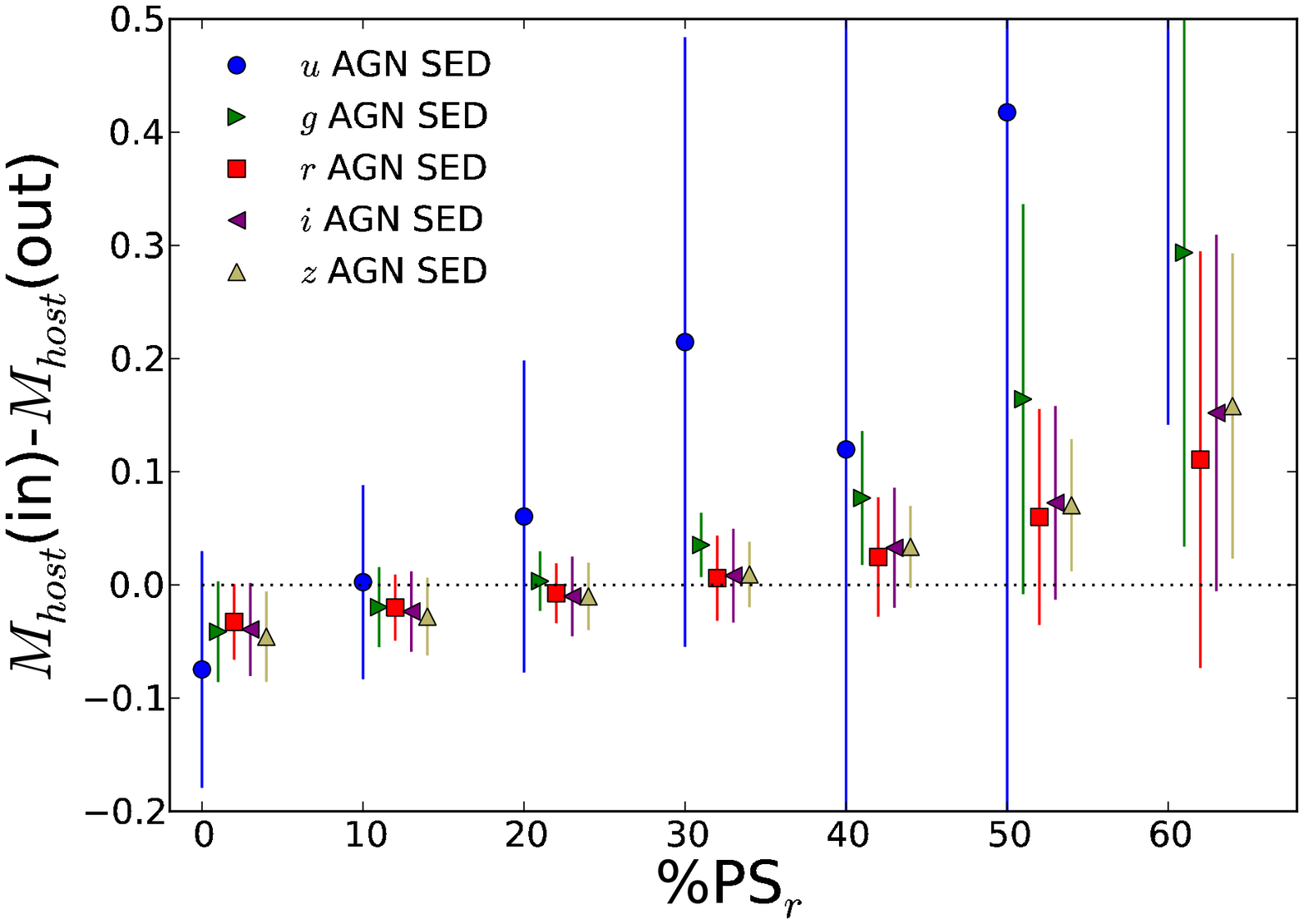} 
\caption{{\em Top}:  Effectiveness of GALFIT in removing the AGN light by filter for modeled AGN galaxies.  We have used a PS and floating S\'{e}rsic Index with n$<$4 to measure the AGN light.   Increasing levels of \%PS are shown.  As the simulated AGN light increases, the galaxy light is overestimated for all filters.  The error bars indicate standard deviations in each bin.  The errors are higher for the $u$ and $z$ where the resolution is poorer.  {\em Bottom}:  We also included the effect of the bluer quasar SED using the colors of AGN \citep{Richards:2006p3527} and the colors of our average Seyfert 2 host galaxy.  The error bars indicate standard deviations in each bin.  The bluer filter performance is worse because of the higher ratio of AGN to galaxy light.}
\label{f16}
\end{figure}

	The next step was to broaden the examination of GALFIT's performance from one filter to the entire \textit{ugriz} filter set.  In Fig.~\ref{f16} (left), we show the performance in each of the filters.  They are similar to each other, but the blue bands have higher uncertainties because of poorer resolution.  

	When testing the performance across filters it is important to consider that the AGN spectral energy distribution (SED) emits more energy at bluer wavelengths than the host galaxy; otherwise we may underestimate the contamination in the bluer bands.  To examine this, we assumed a AGN SED power law of $f_{\nu}^{-0.5}$ as has been found for the optical spectrum of quasars \citep{Richards:2006p3527}.  We then normalized to the total light in each filter band based off of the \%PS$_{r}$.   
	
	These final modeling results suggest important constraints where GALFIT is effective in removing the AGN light (Fig.~\ref{f16}, right). Based on these results, we restrict our \textit{u} band photometry to galaxies with AGN brightnesses of \%PS$_{r}$$<$20 where the contamination is 0.05$\pm$0.15 mag.  For photometry in the other \textit{griz} filters, a less stringent restriction of \%PS$_{r}$$<$40 is sufficient to keep our errors within $\sigma=0.05\pm$0.04 mag.  
		
\begin{figure} 
\centering 
\includegraphics[width=8.1cm]{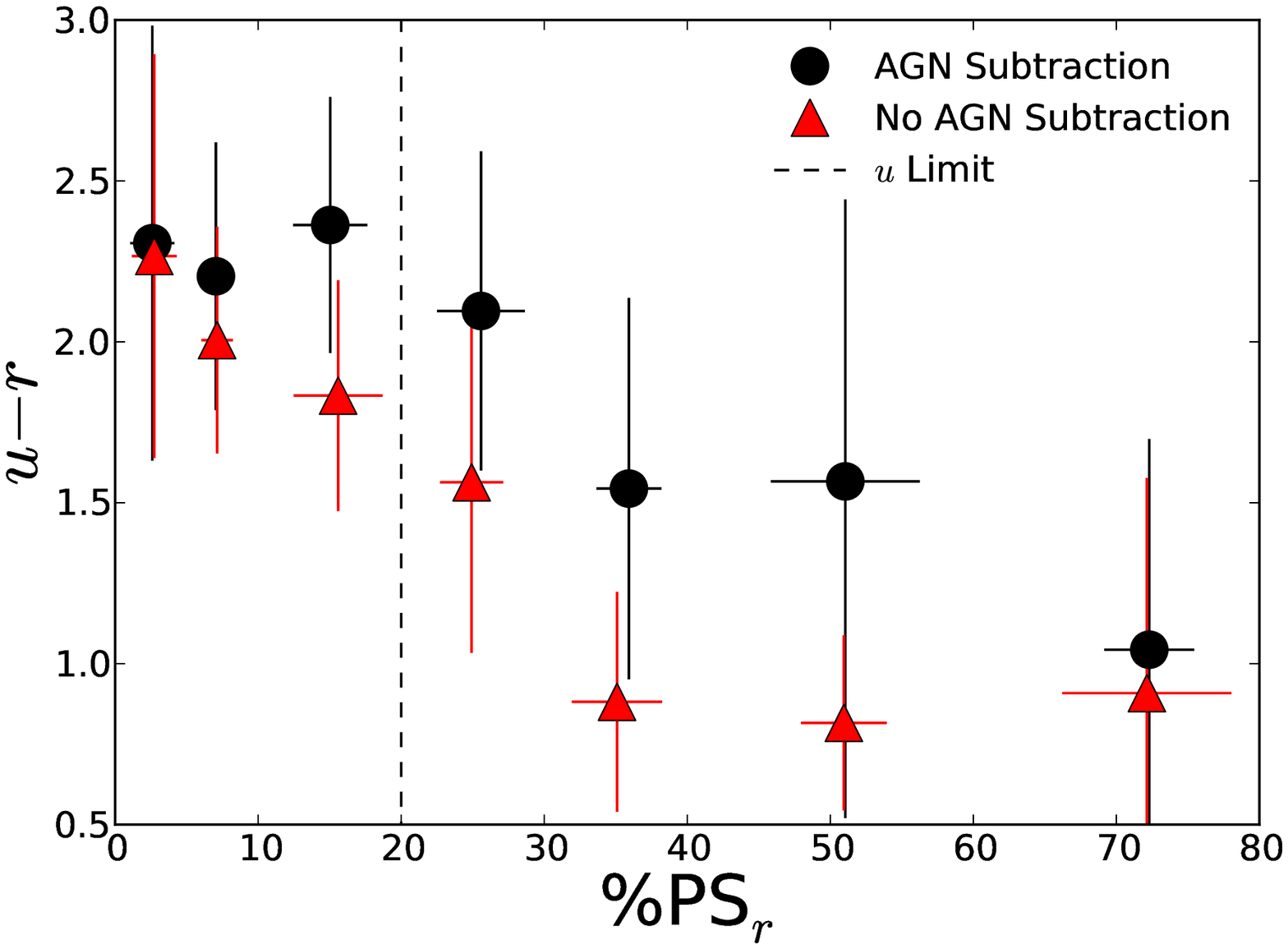} 
\includegraphics[width=8.1cm]{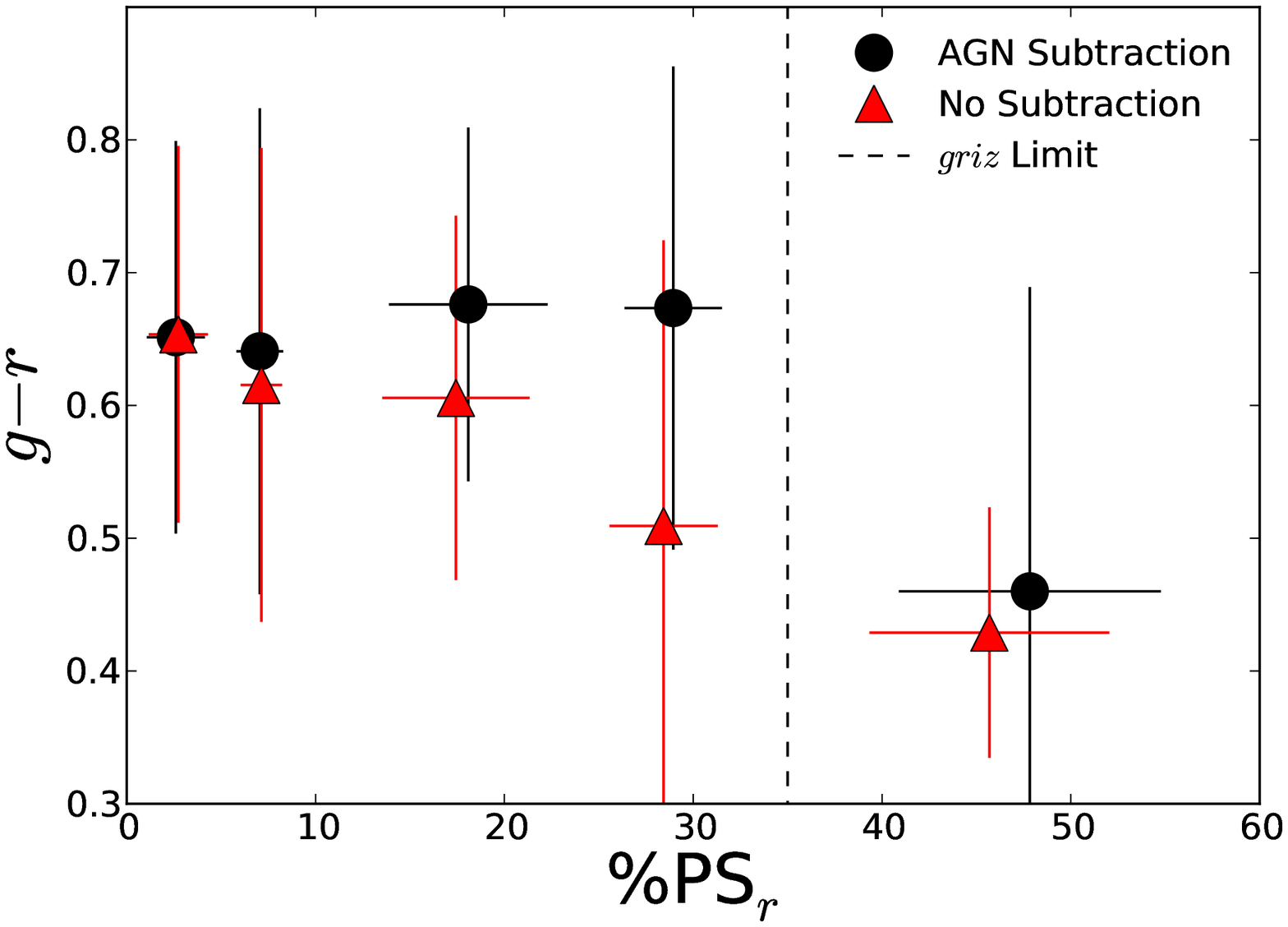} 
\caption{{\em Left}:  Average colors in $u-r$ for BAT AGN before and after AGN subtraction with GALFIT.  The error bars indicate standard deviations in each bin.  Using GALFIT for AGN subtraction,  the colors are almost constant for \%PS$_{r}<$20.  Based on our modeling we believe the strong blueward shift above 20\% is due to AGN contamination not accounted for from GALFIT.  {\em Right}:  Average colors in $g-r$ for BAT AGN before and after AGN subtraction with GALFIT.  The error bars indicate standard deviations in each bin.   The ratio of AGN to galaxy light is smaller in $g$ than $u$ and we see a constant color to \%PS$_{r}<$35.}
\label{f17}
\end{figure}

	In addition to our simulations, we also tested the real BAT AGN galaxies for AGN contamination after removing the AGN light measured by GALFIT.  The effects of subtracting the AGN contribution with GALFIT are shown for the $u-r$ and $g-r$ for the broad-line AGN in Fig.~\ref{f17}.  The galaxy colors stay flat with increasing \% of AGN light up to $\approx$20\% for the \textit{u} band and $\approx$35\% for the  \textit{griz} band.  This agrees with the results of our modeling of AGN contamination at 20\% for the \textit{u} band and 40\% for the  \textit{griz}.  Based on these results, we have imposed a tighter restriction of 35\% AGN light on the  \textit{griz} band photometry to ensure there is no AGN contamination.  

\begin{figure} 
\centering 
\includegraphics[width=8.1cm]{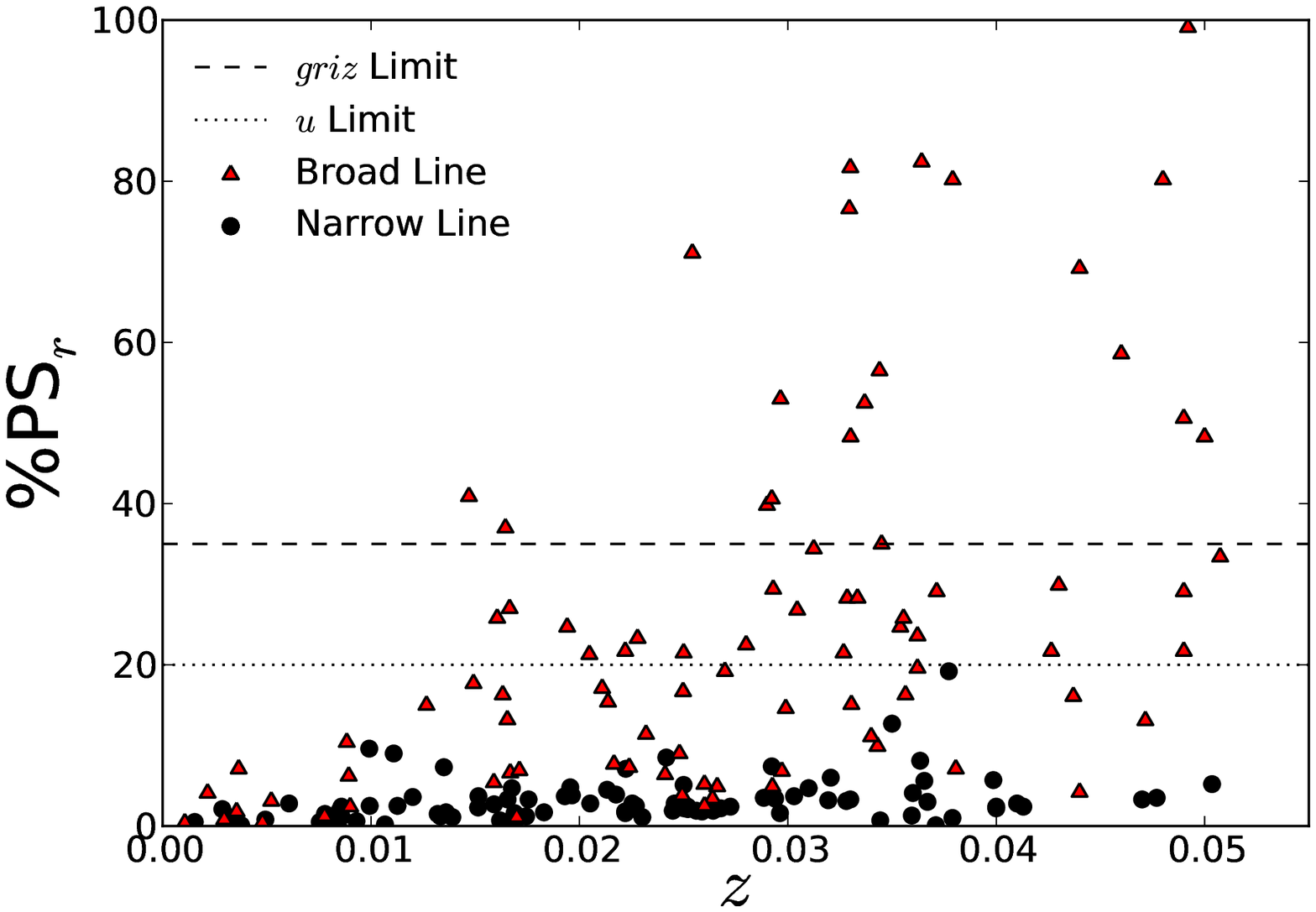} 
\includegraphics[width=8.1cm]{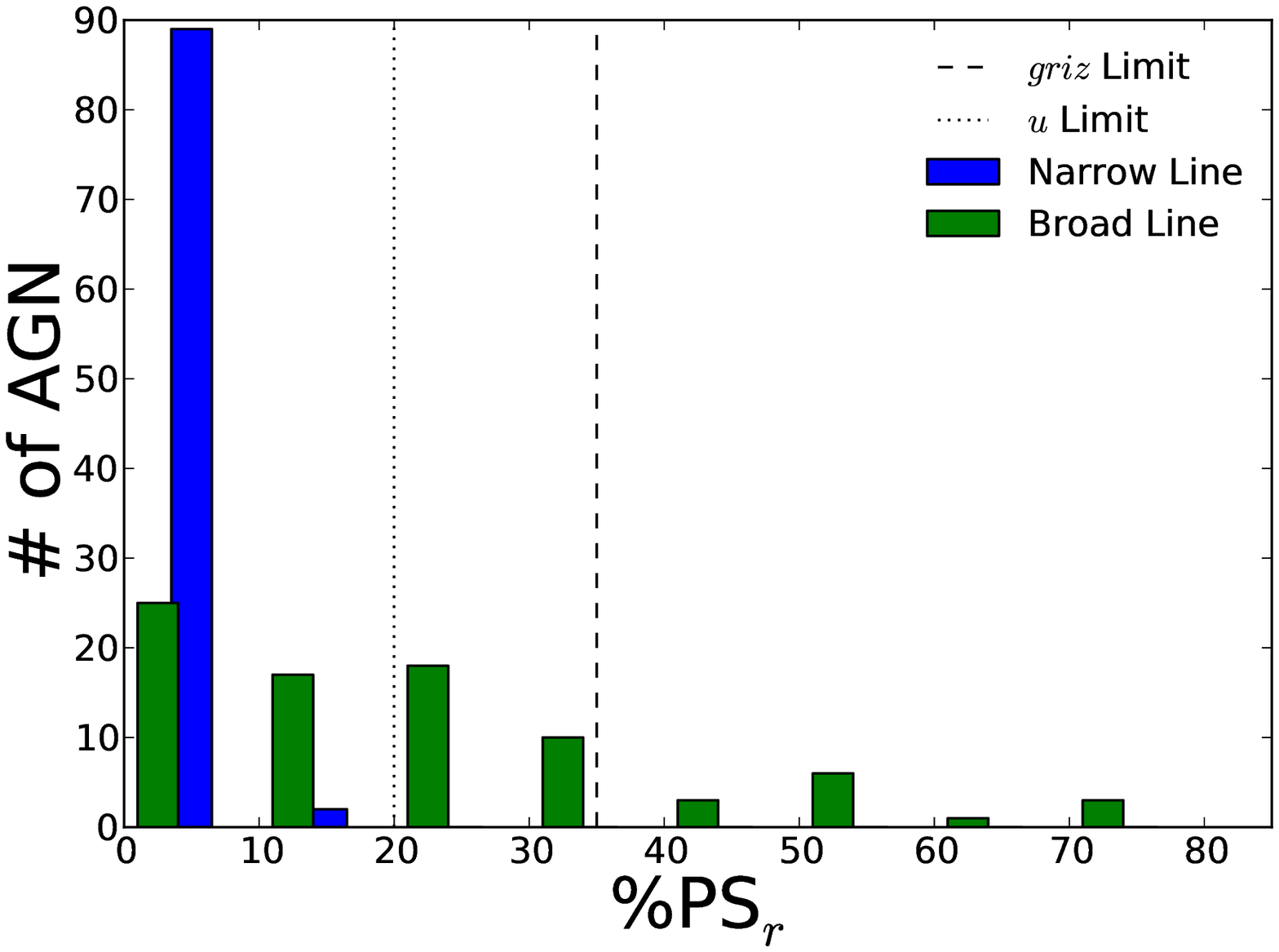} 
\caption{{\em Left}:  Plot of \%PS$_{r}$ for BAT AGN galaxies by redshift.  Narrow-line AGN are represented as dots and broad-line AGN are indicated by triangles. Dashed lines indicate the limits above which GALFIT was unable to effectively remove the AGN light.  These galaxies were not included in the analysis of colors because of AGN contamination.  At higher redshifts we find more BAT AGN with higher ratios of AGN to galaxy light.  {\em Right}:  Plot of number of AGN in bins of \%PS$_{r}$.  We find that GALFIT finds \%PS$_{r}<$5 for most BAT narrow-line BAT AGN.     }
\label{f18}
\end{figure}

	We have restricted our color analysis of host galaxies because of our inability to remove the AGN light with GALFIT for the brightest AGN.  After fitting with GALFIT, 17 or $10\%$ of galaxies had bright broad-line AGN, where \%PS$_{r}$$>$35, and these galaxies were not included in the  \textit{griz} analysis because of uncertainty of the host galaxy photometry (Fig.~\ref{f18}).  When we include both the galaxies that were not included because of pixel saturation and those with with \%PS$_{r}$$>$35, the completeness is 71\% for the highest quartile of BAT luminosity and $>$95\% for the other 3 quartiles.  In the \textit{u} band, 41 galaxies or 23\% were excluded because the luminosity of the AGN exceeded 20$\%$.  The completeness is 71\% for the highest quartile of BAT luminosity and $>$95\% for the other 3 quartiles.  This may introduce a small bias against QSO-like systems. However, in the regions where GALFIT is accurate in the removal of the AGN, we do not see any strong trends towards bluer colors in higher luminosity AGN. 

	Our photometry will be used to provide colors and stellar masses of the host galaxy.  Contamination from the AGN will be reduced in relative color measurements, since the photometry from all filters includes light from the AGN and will to some extent be subtracted off in color measurements.  Finally, when using photometry to determine stellar masses, the redder bands are weighted more heavily, as they tend to be less contaminated by the AGN component than the bluer bands.

\section{Comparison Sample}

% abstract We have performed extensive checks of our optical photometry systematic effects in our sample and have found that the effects of nuclear emission for the Seyfert 1s in our sample can be removed very accurately using GALFIT if they contribute less than 40\% of the total light. Comparison of our photometry and the SDSS shows that for z$<$0.03 a significant fraction of the BAT galaxies have incorrect photometry due to the effects of 'shredding'. 

\begin{figure} 
\includegraphics[width=7cm]{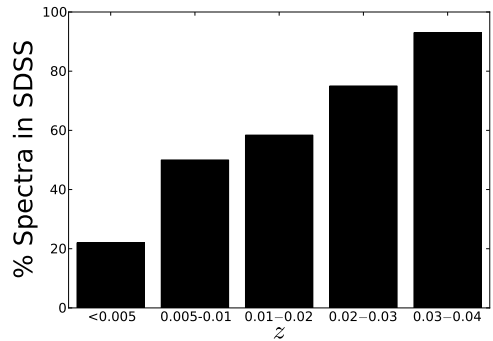}
\centering 
\caption{Percent coverage of BAT AGN in the SDSS spectroscopic footprint with spectroscopy.  Due to the brightness limits (m$_r<15$) in the spectroscopic sample, 30\% of the BAT AGN galaxies are missed by the SDSS spectroscopic sample.  In addition, some broad-line AGN with bright nuclei are misclassified as stars and not included in the SDSS spectroscopy.  Finally, some merging BAT AGN galaxies are not covered because of fiber collision limits in the SDSS.  Due to this effect, we have used SDSS galaxies in the redshift range of 0.01$<$z$<$0.07 to compare to BAT AGN. }
\label{coveragered}
\end{figure}

	In this section, we discuss considerations in our choice of the comparison sample from the SDSS, the use of catalog photometry, and importance of comparing galaxies with similar brightnesses or stellar masses.    We have used galaxies with both photometry and spectroscopy from the SDSS as a comparison sample.  Since 50\% (93/185) of BAT AGN galaxies are in the SDSS spectroscopic coverage area, we can examine these galaxies to determine the completeness of the SDSS catalog sample.   In Fig.~\ref{coveragered}, the SDSS spectroscopic coverage of BAT AGN by redshift bin is shown.  While 70\% of the BAT sample in the SDSS has spectroscopic coverage, the brightest galaxies in each redshift bin are missed for z$<$0.03.  Above this redshift a few broad-line AGN with bright nuclei are misclassified as stars.  Due to the selection effect against the brightest galaxies, in this study we have chosen a SDSS control sample of galaxies with 0.01$<$z$<$0.07. 

\begin{figure} 
\centering 
\plottwo{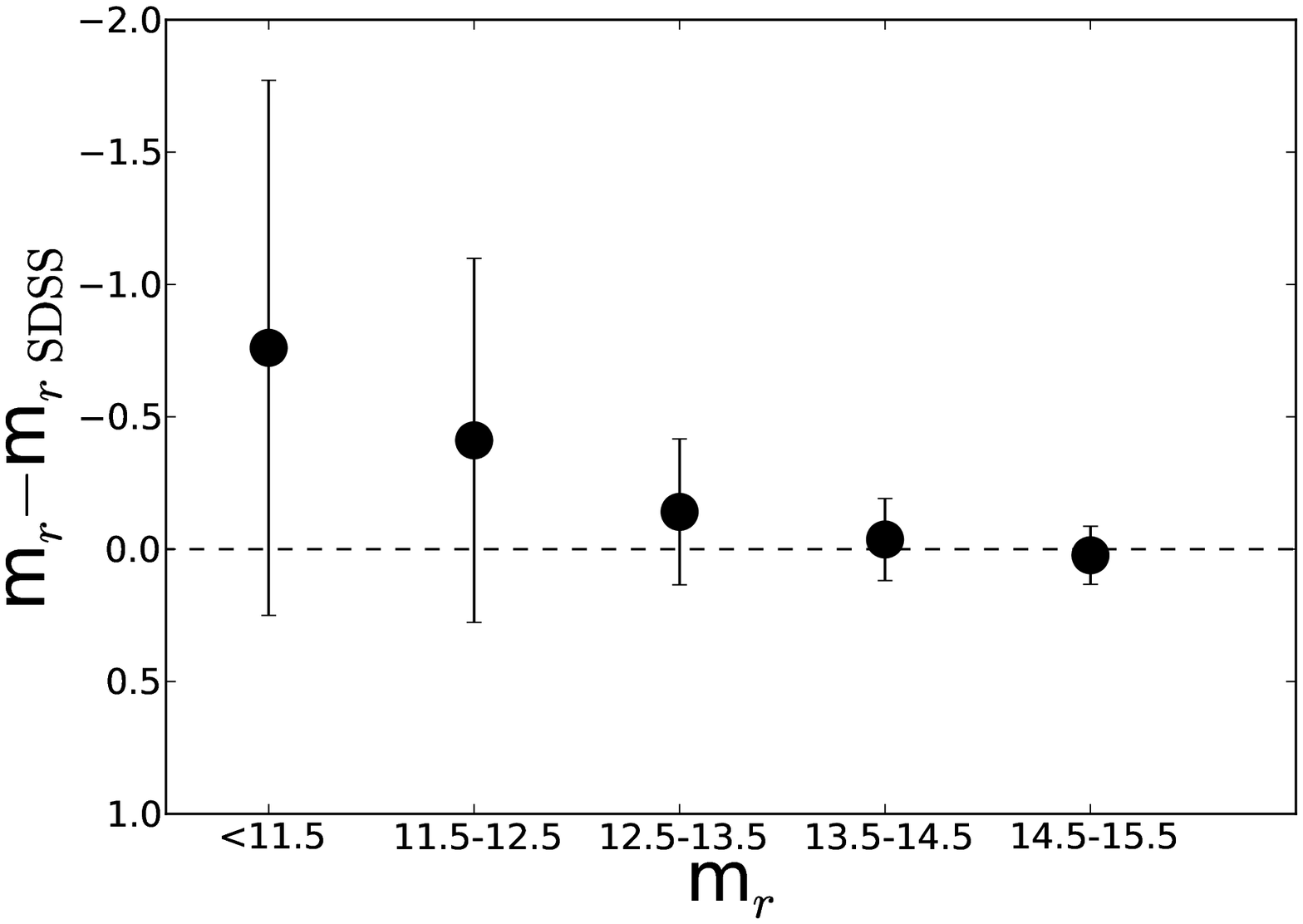}{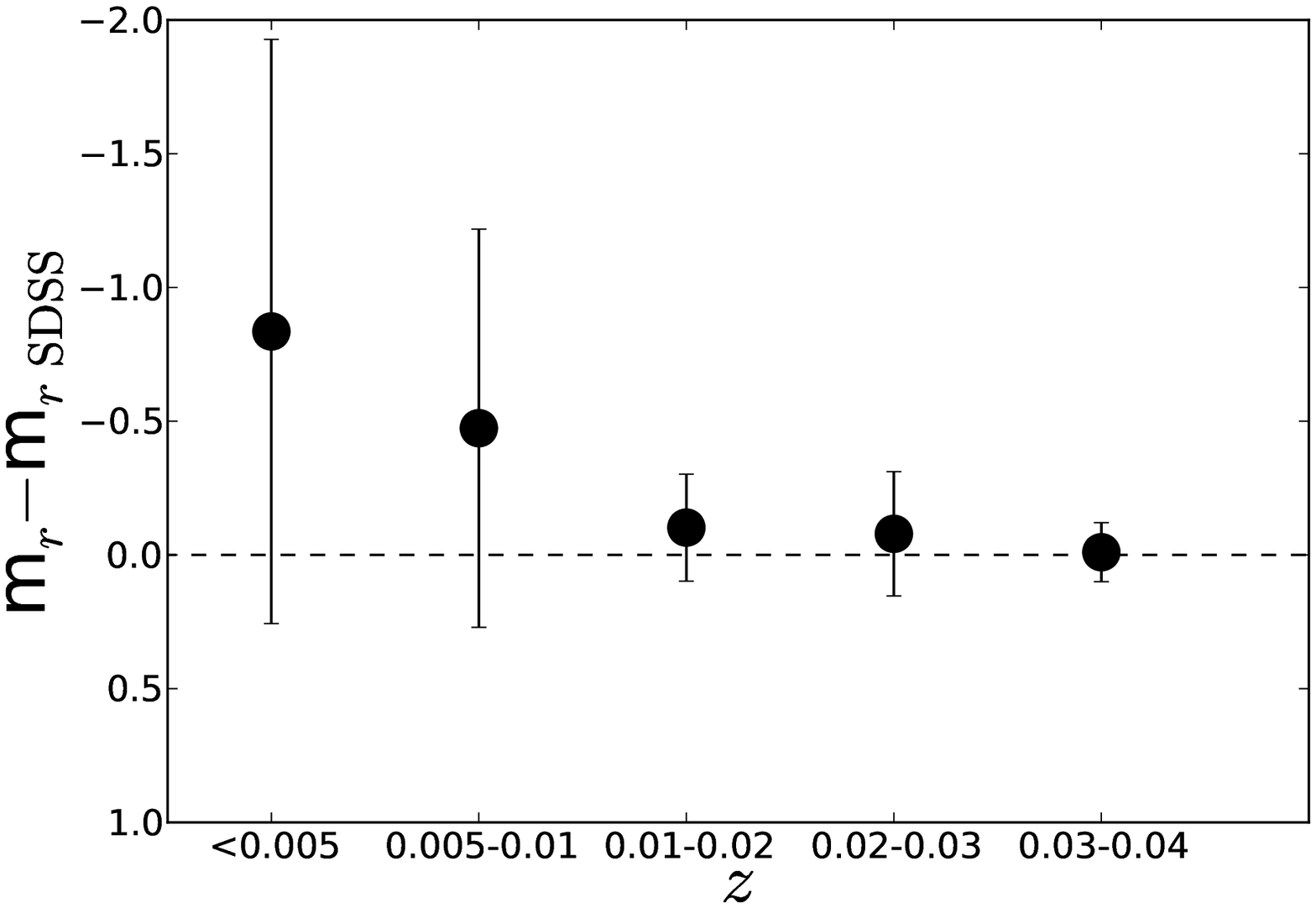} 
\caption{Comparison of our measured Petrosian magnitudes to the SDSS catalog values for a range of magnitudes (m$_{r}$) ({\em left}) and redshifts ({\em right}).   Negative values indicate we find a brighter magnitude than the SDSS catalog.  Since the automated routine in the SDSS has a tendency to shred galaxies into multiple component galaxies, the magnitudes are reduced.  This shredding effect is much stronger for the brightest galaxies.  However, for m$_{r}$$>$$13.5$ or z$>$0.01 the magnitudes are in good agreement.  Given these results, we have restricted our SDSS catalog comparison to this range. }
\label{petrocomp}
\end{figure}

\begin{figure} 
\centering 
\includegraphics[width=8.1cm]{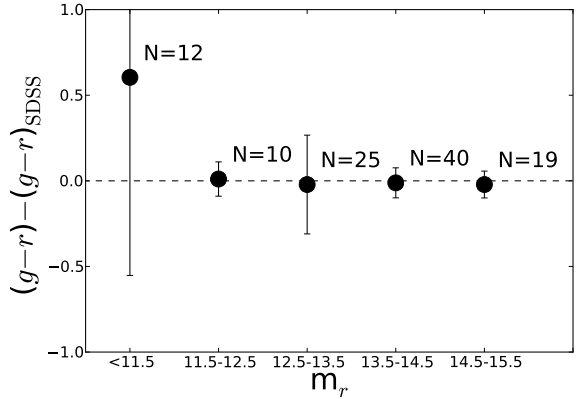}
\caption{Comparison of the average difference in $g-r$ color between our measurements and the SDSS catalog. The error bars indicate standard deviations in each bin.  The SDSS catalog colors show better agreement than apparent magnitudes since the effects of shredding tend to cancel each other out in the reduced brightness of both filters.}
\label{colocomp}
\end{figure}

	We also ensured that  our own photometry of BAT AGN agreed with the SDSS catalog measurements.  We can measure the differences between our photometry and those in the SDSS catalog since 62\% (116/185) of BAT AGN are in the SDSS photometric catalog (Fig.~\ref{petrocomp}).   The SDSS photometric catalog incorrectly shreds features of bright, nearby galaxies, such as spiral arms, rings, and dust lanes into different components.  This causes a systematic underestimation of the brightness of galaxies and variations in their measured color.   Because of this shredding effect, we have restricted our SDSS catalog comparison to galaxies with m$_{r}$$>$$13.5$ and z$>$0.01, where we find good agreement in photometry.  In Fig.~\ref{colocomp},  we show a comparison between the $g-r$ of BAT AGN measured in the SDSS catalog and measured using our own photometry.  The SDSS catalog colors show better agreement than the photometry since the effects of shredding tend to cancel each other out in relative color measurements. 
	
\begin{figure} 
\centering 
\includegraphics[width=8.1cm]{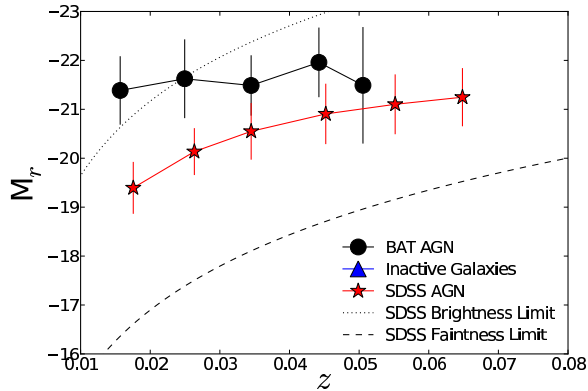} 
\caption{Plot of average absolute mag in $r$ for the BAT AGN, SDSS AGN, and inactive galaxies.  The error bars indicate standard deviations in each bin.  The dashed lines indicate the approximate brightness and faintness limits from the SDSS spectroscopic survey.  At higher redshifts, the SDSS detects galaxies that are more luminous and have a higher stellar mass because of the selection effect against faint galaxies.  On the other hand, the BAT AGN survey detects AGN galaxies of a constant optical brightness across a range of redshifts.  Due to the selection effects it is important to compare host galaxy colors between the BAT survey and SDSS survey only at similar brightnesses or stellar masses.  For comparison, in the redshift range between 0.03 to 0.05 and the survey coverage area of the SDSS, the BAT survey has 28 broad-line and 17 NL AGN.  In this same range, the SDSS has 121 broad-line and 411 Seyfert 2 AGN. }
\label{rmags}
\end{figure}

	When comparing host galaxy properties, it is important to consider the flux limited nature of both the SDSS and BAT surveys.  In Fig.~\ref{rmags}, a plot of average M$_r$ by redshift  for the BAT AGN, SDSS AGN, and inactive SDSS galaxies is shown. At higher redshifts the SDSS detects AGN and galaxies that are more luminous and have a higher stellar mass because of a selection effect against optically faint galaxies.  On the other hand, the BAT AGN survey detects AGN galaxies of a constant optical brightness across a range of redshifts.  Due to these selection effects, it is important to compare host galaxy colors between the BAT survey and SDSS survey only at similar brightnesses or stellar masses. 

\section{Selection Effects in the BAT Survey}
		
\begin{figure} 
\centering 
\includegraphics[width=8.1cm]{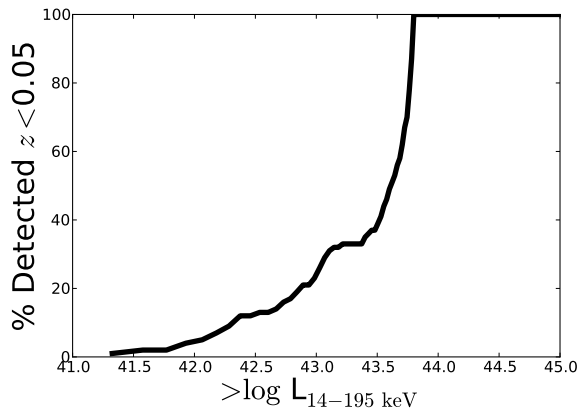} 
\caption{Plot of estimated completeness above a ultra hard X-ray luminosity within the redshift range of this survey (z$<$0.05) using the median flux sensitivity of the 58 month survey ($1.1\times10^{-11}$ erg cm$^{-2}$ s$^{-1}$; Baumgartner et al.~2011, submitted).  This plot assumes the ultra hard X-ray AGN are randomly distributed by volume and uses the distribution of lower redshift sources to estimate those missed at higher redshift.  We find that the BAT sources are complete for z$<0.05$ in this survey for log L$_{14-195\,\rm{keV}}>43.7$ or $\approx$log L$_{2-10\,\rm{keV}}>$43.2 assuming no intrinsic absorption.  }
\label{xraycomp}
\end{figure}

\begin{figure} 
\centering 
\includegraphics[width=8.1cm]{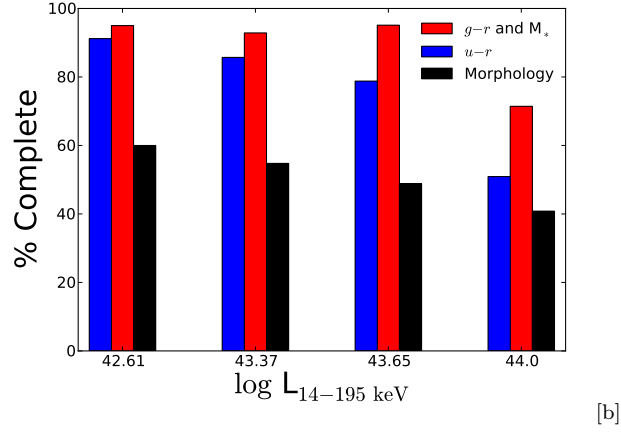}[b] 
\caption{Completeness of various measurements at each quartile of BAT luminosity. Morphology measurements were limited to narrow-line AGN while color measurements exclude systems with very
bright AGN.  The highest quartile of BAT luminosity is less complete than the lowest quartile, although this difference is $<$20\%.   
}
\label{batlims}
\end{figure}

	In the ultra hard X-rays, the BAT survey is also flux limited.  Assuming the ultra hard X-ray AGN are distributed randomly following the distribution of luminosities at lower redshifts, and using the limiting sky sensitivity, we can make a further estimate of completeness (Fig.~\ref{xraycomp}).  We find that the BAT sources are complete for z$<$$0.05$ in this survey for log L$_{14-195\,\rm{keV}}$$>$43.7 or $\approx$log L$_{2-10\,\rm{keV}}$$>$43.2 assuming no intrinsic absorption.   In addition, we limited our analysis of morphologies to NL BAT AGN.  The completeness fractions are shown in Fig.~\ref{batlims} as a function of ultra hard X-ray luminosity.  We see that the highest luminosity quartile for BAT luminosity is less complete than the lowest quartile, although this difference is $<$20\%.   

\begin{figure} 
\centering 
\includegraphics[width=8.1cm]{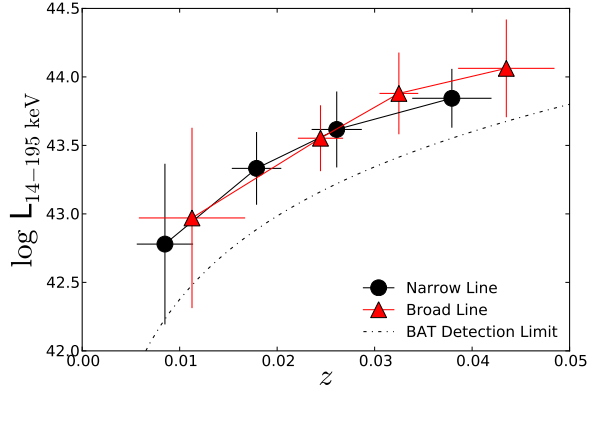}[b] 
\caption{Plot of average ultra hard X-ray luminosity compared to redshift for narrow-line and broad-line AGN in BAT.   The error bars indicate standard deviations in each bin. The dashed line shows the approximate flux limit of the BAT survey.  This shows that the  BAT survey does have a slight tendency to find narrow-line AGN at closer redshifts than broad-line AGN.   The mean redshift is 0.027 for broad-line sources and 0.022 for narrow-line sources.  The mean log L$_{14-195\,\rm{keV}}$ for broad-line sources is 43.56$\pm$0.65 and 43.37$\pm$0.59. }
\label{batlumsplit}
\end{figure}

	  In addition, the BAT survey may miss heavily obscured Compton-thick sources that may be identified using methods at other wavelengths.  Compton-thick sources are AGN where our line of sight to the source is blocked with obscuring matter that has an optical depth of $\tau>$ 1 (N$_{\rm H}>1.5\times10^{24}$ cm$^{-2}$ ).  At these optical depths, much of the X-ray emission is reflected and not direct. For Compton-thick sources, the column densities are so high that little to no direct emission escapes below 10 keV.  Estimates of the number of Compton-thick sources in the BAT AGN sample have ranged from 3\% to 20\% \citep{Tueller:2008p2710,Winter:2009p79,Ajello:2008p9070}.  A recent analysis of the \textit{INTEGRAL} AGN at 20--40 keV finds that the number of Compton-Thick AGN found by optical and ultra hard X-ray methods is in agreement up to z$=$0.015 \citep{Malizia:2009p8784}.

	We independently estimated the number of missing Compton-thick sources by investigating the difference in narrow and broad-line sources by redshift.  A plot of average ultra hard X-ray luminosity compared to redshift for narrow-line and broad-line AGN in the BAT survey can be found in Fig.~\ref{batlumsplit}.  We have also plotted the approximate all-sky limiting flux of the BAT ultra hard X-ray detections for the 58 month catalog.   This shows that the  BAT survey does have a slight tendency to find narrow-line AGN at closer redshifts than broad-line AGN.   The mean redshift is 0.027 for broad-line sources and 0.022 for narrow-line sources.  The mean $\log$ L$_{14-195\,\rm{keV}}$ is 43.56$\pm$0.65 for broad-line sources and is 43.37$\pm$0.59 for the narrow-line sources.

\begin{figure*} 
\centering 
\includegraphics[width=8.1cm]{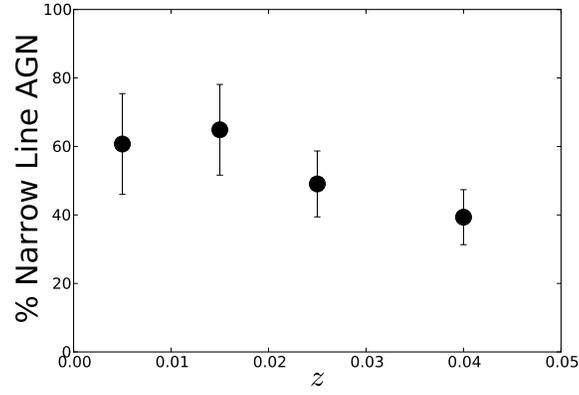} 
\caption{Percent of narrow-line sources in different redshift ranges in the BAT AGN sample.  The error bars represent 1$\sigma$ Poisson statistics. We find that the number of narrow-line sources falls at higher redshifts.  For z$<$0.01, 61\% are NL AGN while at 0.03$<$z$<$0.05 only 31\% are NL AGN.  If we assume that the ratio of 61\% narrow-line AGN is constant across redshift, we will be missing about 50 narrow-line sources at higher redshift or 27\% of the entire sample.}
\label{narrowbyz}
\end{figure*}

	An additional way to estimate the number of missed absorbed sources is by measuring the percentage of NL BAT AGN by redshift (Fig.~\ref{narrowbyz}).  We find that the number of narrow-line sources falls at higher redshifts.  For z$<$0.01, 61\% are NL AGN while at 0.03$<$z$<$0.05 only 31\% are NL AGN.  If we assume that the ratio of 61\% narrow-line AGN in the z$<$0.01 bin is the true value and is constant with redshift, then we will be missing about 50 narrow-line sources at higher redshift or 27\% of the entire sample.  However, we do not find any statistically significant difference in color between NL and broad-line AGN or between luminous ($\log$ L$_{14-195\,\rm{keV}}$$>$43.5)  and less luminous sources.  We also do not find any statistically significant difference with increasing X-ray column densities.  These results suggest that the flux-limited nature of the survey does not strongly influence our overall results.
	
	We can make an additional estimate of the number of missing Compton-thick AGN based on the ratio of NL to broad-line AGN in the SDSS survey.  In the redshift range between 0.03 to 0.05, the SDSS has 121 broad-line AGN and 411 NL AGN.  For comparison, the BAT survey has 28 broad-line AGN and 17 NL AGN in this range.  This suggests that approximately 77\% (411/532) are narrow-line, which is greater than the 61\% of NL BAT AGN at low redshift, but not outside of the 1 sigma error bars for z$<$0.02.  We may therefore estimate that at a maximum $\approx$16\% AGN are missed as Compton-thick.  Unless a  large fraction of the missed sources have systematically different colors, morphologies or mass than the detected sources we do not expect a large effect on our results.  	
	
\clearpage

\begin{center}
\begin{longtable}{l l l l l l l l} 
%\tabletypesize{\scriptsize}
%\tablecolumns{8} 
%\tablewidth{0pc} 
\caption[Summary of Kitt Peak Observations]{Summary of Kitt Peak Observations}  \label{allagnkp} \\
\hline \hline \\[-2ex]
   \multicolumn{1}{c}{\textbf{Name}} &
   \multicolumn{1}{c}{\textbf{Date Obs.}} &
   \multicolumn{1}{c}{\textbf{Type \tablenotemark{1}}} &
   \multicolumn{1}{c}{\textbf{$z$}} &
   \multicolumn{1}{c}{\textbf{Dist. (Mpc)\tablenotemark{2}}} &
   \multicolumn{1}{c}{\textbf{E(B-V)\tablenotemark{3}  }} &
   \multicolumn{1}{c}{\textbf{Airmass\tablenotemark{4}}} &
   \multicolumn{1}{c}{\textbf{PSF($\arcsec$) \tablenotemark{5}}} 
\\[0.5ex] \hline
   \\[-1.8ex]
\endfirsthead

\multicolumn{3}{c}{{\tablename} \thetable{} -- Continued} \\[0.5ex]
   \multicolumn{1}{c}{\textbf{Name}} &
   \multicolumn{1}{c}{\textbf{Date Obs.}} &
   \multicolumn{1}{c}{\textbf{Type \tablenotemark{1}}} &
   \multicolumn{1}{c}{\textbf{$z$}} &
   \multicolumn{1}{c}{\textbf{Dist. (Mpc)\tablenotemark{2}}} &
   \multicolumn{1}{c}{\textbf{E(B-V)\tablenotemark{3}  }} &
   \multicolumn{1}{c}{\textbf{Airmass\tablenotemark{4}}} &
   \multicolumn{1}{c}{\textbf{PSF($\arcsec$) \tablenotemark{5}}} 
\\[0.5ex] \hline
   \\[-1.8ex]

\endhead

\multicolumn{3}{l}{{Continued on Next Page\ldots}} \\
\endfoot
%This is the footer for the last page of the table...
  \\[-1.8ex] \hline \hline
\endlastfoot
1RXS J045205.0+493248 & 2008-02-07 & Sy1  & 0.029 & 125.3 & 0.72 & 1.1 & 1.2\\
2MASX J00253292+6821442 & 2008-11-06 & Sy2 & 0.012 & 51.2 & 1.04 & 1.2 & 2.5\\
2MASX J04440903+2813003 & 2008-02-08 & Sy2 & 0.011 & 48 & 0.85 & 1.1 & 1.0\\
2MASX J05054575-2351139 & 2008-11-06 & Sy2 & 0.035 & 151.9 & 0.03 & 1.7 & 1.5\\
2MASX J06411806+3249313 & 2008-02-09 & Sy2 & 0.047 & 205.8 & 0.15 & 1.0 & 1.0\\
2MASX J09043699+5536025 & 2008-02-13 & Sy1  & 0.037 & 161.4 & 0.02 & 1.1 & 1.2\\
2MASX J09112999+4528060 & 2008-02-12 & Sy2  & 0.026 & 115.5 & 0.01 & 1.0 & 1.4\\
2MASX J11454045-1827149 & 2008-02-06 & Sy1  & 0.032 & 142.7 & 0.03 & 1.6 & 1.4\\
\end{longtable}
\end{center}

\clearpage

\begin{center}
\begin{longtable}{l l l l l l l l} 
%\tabletypesize{\scriptsize}
%\tablecolumns{8} 
%\tablewidth{0pc} 
\caption[Summary of SDSS Observations]{Summary of SDSS Observations}  \label{allagnsdss} \\
\hline \hline \\[-2ex]
   \multicolumn{1}{c}{\textbf{Name}} &
   \multicolumn{1}{c}{\textbf{Date Obs.}} &
   \multicolumn{1}{c}{\textbf{Type \tablenotemark{1}}} &
   \multicolumn{1}{c}{\textbf{$z$}} &
   \multicolumn{1}{c}{\textbf{Dist. (Mpc)\tablenotemark{2}}} &
   \multicolumn{1}{c}{\textbf{E(B-V)\tablenotemark{3}  }} &
   \multicolumn{1}{c}{\textbf{Airmass\tablenotemark{4}}} &
   \multicolumn{1}{c}{\textbf{PSF($\arcsec$) \tablenotemark{5}}} 
\\[0.5ex] \hline
   \\[-1.8ex]
\endfirsthead

\multicolumn{3}{c}{{\tablename} \thetable{} -- Continued} \\[0.5ex]
   \multicolumn{1}{c}{\textbf{Name}} &
   \multicolumn{1}{c}{\textbf{Date Obs.}} &
   \multicolumn{1}{c}{\textbf{Type \tablenotemark{1}}} &
   \multicolumn{1}{c}{\textbf{$z$}} &
   \multicolumn{1}{c}{\textbf{Dist. (Mpc)\tablenotemark{2}}} &
   \multicolumn{1}{c}{\textbf{E(B-V)\tablenotemark{3}  }} &
   \multicolumn{1}{c}{\textbf{Airmass\tablenotemark{4}}} &
   \multicolumn{1}{c}{\textbf{PSF($\arcsec$) \tablenotemark{5}}} 
\\[0.5ex] \hline
   \\[-1.8ex]

\endhead

\multicolumn{3}{l}{{Continued on Next Page\ldots}} \\
\endfoot
%This is the footer for the last page of the table...
  \\[-1.8ex] \hline \hline
\endlastfoot

2MASX J01064523+0638015 & 2005-09-28 & Sy2 & 0.041 & 178.7 & 0.05 & 1.1 & 1.0\\
2MASX J03305218+0538253 & 2004-12-14 & Sy1 & 0.046 & 201.3 & 0.37 & 1.1 & 1.7\\
2MASX J03534246+3714077 & 2003-01-28 & Sy2 & 0.018 & 78.4 & 0.53 & 1.0 & 0.9\\
2MASX J07595347+2323241 & 2003-01-25 & Sy2  & 0.029 & 128 & 0.05 & 1.4 & 1.4\\
2MASX J16174158+0607100 & 2003-04-25 & Sy2 & 0.037 & 164.8 & 0.08 & 1.1 & 0.9\\
ARP 151 & 2002-04-01 & Sy1 & 0.021 & 90.5 & 0.02 & 1.2 & 1.5\\
CGCG 031-072 & 2002-03-05 & Sy1 & 0.033 & 143.2 & 0.03 & 1.2 & 1.8\\
CGCG 046-033 & 2001-04-16 & Sy1.5 & 0.034 & 148.7 & 0.03 & 1.2 & 1.2\\
\end{longtable}
\end{center}

\clearpage

\begin{deluxetable}{l l l l l l l l l} 
\tablecolumns{12} 
\tablewidth{0pc} 
\tablecaption{Observing Summary--Filter Limiting Magnitudes and PSF} 
\tablehead{
\colhead{} &\colhead{$u$}&\colhead{$g$} & \colhead{$r$}&\colhead{$i$}&\colhead{$z$}&
}
\startdata
SDSS  $[mag]$\tablenotemark{1}   &22.3     &23.3     &22.9&22.3&20.7\\
Kitt Peak $[mag]$ &24.2	    &24.8	& 24.4&23.7&22.0\\
\hline
SDSS (\arcsec)\tablenotemark{2}     &1.36$\pm$0.22&1.30$\pm$0.19&1.17$\pm$0.17&1.14$\pm$0.18&1.15$\pm$0.18\\
Kitt Peak (\arcsec) &1.67$\pm$0.76&1.56$\pm$0.47&1.46$\pm$0.41&1.44$\pm$0.59&1.42$\pm$0.48
\enddata
\tablenotetext{1}{Detection based on overlapping fields in SDSS and Kitt Peak sample with at least 5 pixels above signal-to-noise ratio$>$3.}
\tablenotetext{2}{PSF function was calculated using the FWHM of a single gaussian fit to the 10 brightest unsaturated stars.}
\label{limmagcomp}
\end{deluxetable}

$^{}$
\vspace{3 mm}

\begin{center}
\begin{longtable}{l c c c c c c c c c c c}
%\tabletypesize{\scriptsize}
%\tablecolumns{8} 
%\tablewidth{0pc} 
\caption[Summary of Photometry and Derived Properties]{Summary of Photometry and Derived Properties}  \label{tablephot} \\
\hline \hline \\[-2ex]
   \multicolumn{1}{c}{\textbf{Galaxy}} &
   \multicolumn{1}{c}{\textbf{$u$\tablenotemark{1}}} &
   \multicolumn{1}{c}{\textbf{$g$}} &
   \multicolumn{1}{c}{\textbf{$r$}} &
   \multicolumn{1}{c}{\textbf{$i$}} &
   \multicolumn{1}{c}{\textbf{$z$}} &
   \multicolumn{1}{c}{\textbf{$\log$ M$_*$\tablenotemark{2}}} &
   \multicolumn{1}{c}{\textbf{PS$_r$\tablenotemark{3}   }} &
   \multicolumn{1}{c}{\textbf{E($g-r$)\tablenotemark{4}}} &
   \multicolumn{1}{c}{\textbf{E($g-r$) Incl. \tablenotemark{5}}} &
   \multicolumn{1}{c}{\textbf{$u$\tablenotemark{6}}} &
   \multicolumn{1}{c}{\textbf{$griz$\tablenotemark{7}}} \\

   \multicolumn{1}{c}{} &
   \multicolumn{1}{c}{$mag$} &
   \multicolumn{1}{c}{$mag$} &
   \multicolumn{1}{c}{$mag$} &
   \multicolumn{1}{c}{$mag$} &
   \multicolumn{1}{c}{$mag$} &
   \multicolumn{1}{c}{M$_\odot$} &
   \multicolumn{1}{c}{\%} &
   \multicolumn{1}{c}{$mag$} &
   \multicolumn{1}{c}{$mag$} &
   \multicolumn{1}{c}{Contam.} &
   \multicolumn{1}{c}{Contam.} 
        
\\[0.5ex] \hline
   \\[-1.8ex]
\endfirsthead

\multicolumn{3}{c}{{\tablename} \thetable{} -- Continued} \\[0.5ex]
    \multicolumn{1}{c}{\textbf{Galaxy}} &
   \multicolumn{1}{c}{\textbf{$u$\tablenotemark{1}}} &
   \multicolumn{1}{c}{\textbf{$g$}} &
   \multicolumn{1}{c}{\textbf{$r$}} &
   \multicolumn{1}{c}{\textbf{$i$}} &
   \multicolumn{1}{c}{\textbf{$z$}} &
   \multicolumn{1}{c}{\textbf{$\log$ M$_*$\tablenotemark{2}}} &
   \multicolumn{1}{c}{\textbf{PS$_r$\tablenotemark{3}   }} &
   \multicolumn{1}{c}{\textbf{E($g-r$)\tablenotemark{4}}} &
   \multicolumn{1}{c}{\textbf{E($g-r$) Incl. \tablenotemark{5}}} &
   \multicolumn{1}{c}{\textbf{$u$\tablenotemark{6}}} &
   \multicolumn{1}{c}{\textbf{$griz$\tablenotemark{7}}} \\

   \multicolumn{1}{c}{} &
   \multicolumn{1}{c}{$mag$} &
   \multicolumn{1}{c}{$mag$} &
   \multicolumn{1}{c}{$mag$} &
   \multicolumn{1}{c}{$mag$} &
   \multicolumn{1}{c}{$mag$} &
   \multicolumn{1}{c}{M$_\odot$} &
   \multicolumn{1}{c}{\%} &
   \multicolumn{1}{c}{$mag$} &
   \multicolumn{1}{c}{$mag$} &
   \multicolumn{1}{c}{Contam.} &
   \multicolumn{1}{c}{Contam.}    
\\[0.5ex] \hline
   \\[-1.8ex]

\endhead

\multicolumn{3}{l}{{Continued on Next Page\ldots}} \\
\endfoot
%This is the footer for the last page of the table...
  \\[-1.8ex] \hline \hline
\endlastfoot

2MASX J00253292+6821442 & $\cdot \cdot \cdot$ & 12.86 & 12.14 & 11.85 & 11.9 & 10.05 & 3.59 & 0.03&$\cdot \cdot \cdot$& & \\
2MASX J01064523+0638015 & 16.9 & 15.54 & 14.76 & 14.27 & 14.15 & 10.4 & 2.75 & 0.14&0.01& & \\
2MASX J03534246+3714077 & 15.84 & 13.99 & 13.5 & 13.31 & 13.18 & 9.9 & 1.67 & 0.21&$\cdot \cdot \cdot$& & \\
2MASX J04440903+2813003 & $\cdot \cdot \cdot$ & 12.68 & 11.8 & 11.4 & 11.3 & 10.41 & 2.51 & 0.24&$\cdot \cdot \cdot$& & \\
2MASX J05054575-2351139 & 17.31 & 16.11 & 15.44 & 15.12 & 15.36 & 9.13 & 12.7 & 0.03&$\cdot \cdot \cdot$& & \\
2MASX J06411806+3249313 & 17.9 & 16.56 & 15.95 & 15.56 & 15.25 & 9.94 & 3.25 & 0.44&$\cdot \cdot \cdot$& & \\
2MASX J07595347+2323241 & 15.44 & 14.11 & 13.41 & 13.02 & 12.65 & 10.57 & 1.59 & 0.34&0.04& & \\
2MASX J09043699+5536025 & 17.61 & 16.25 & 15.81 & 15.31 & 15.16 & 9.76 & 29.1 & 0.62&$\cdot \cdot \cdot$&Y & \\
\end{longtable}
\end{center}

\clearpage

\begin{deluxetable}{l l l | l l | l l l} 
\tablecolumns{7} 
\tablewidth{0pc} 
\tablecaption{Summary of Color Comparison} 
%\addtolength{\tabcolsep}{-2pt}
\tabletypesize{\scriptsize}
\tablehead{
\colhead{$\log$ M$_*$} & \colhead{Type} & \colhead{N} & \colhead{Mean ($g-r$)} &\colhead{P(K-S)\%\tablenotemark{1}}& \colhead{Mean ($u-r$)}& \colhead{P(K-S)\%}}
\startdata
9.6 to 10     &BAT&37&0.62$\pm$0.15&&2.42$\pm$0.91\\
                     &Inactive&23041&0.63$\pm$0.15&41&2.03$\pm$0.43&10\\
                     &SDSS AGN&455&0.72$\pm$0.10&$<$0.01&2.31$\pm$0.29&36\\
\hline
10 to 10.3   &BAT&39&0.68$\pm$0.12&&2.18$\pm$0.36&\\
                     &Inactive&7542&0.76$\pm$0.12&$<$0.01&2.40$\pm$0.38&$<$0.01\\
                     &SDSS AGN&743&0.76$\pm$0.09&$<$0.01&2.40$\pm$0.39&$<$0.01\\
\hline
10.3 to 10.6&BAT&39&0.69$\pm$0.13&&2.26$\pm$0.57&\\
                     &Inactive&3108&0.81$\pm$0.12&$<$0.01&2.56$\pm$0.22&$<$0.01\\
                     &SDSS AGN&397&0.78$\pm$0.09&$<$0.01&2.48$\pm$0.28&$<$0.01\\
\hline
$>$10.6       &BAT&39&0.70$\pm$0.14&&2.23$\pm$0.67&\\
                      &Inactive&820&0.81$\pm$0.04&$<$0.01&2.78$\pm$0.20&$<$0.01\\
                     &SDSS AGN&148&0.83$\pm$0.07&$<$0.01&2.65$\pm$0.27&$<$0.01\\
\enddata
\tablenotetext{1}{Probability percentage from a Kolmogorov-Smirnov (K-S) test that the BAT colors are from the same parent population.}
\label{tabcolor}
\end{deluxetable}

$^{}$
\vspace{3 mm}

\begin{center}
\begin{longtable}{l l l l l l} 
\caption[Summary of Morphological Properties]{Summary of Morphological Properties}  \label{tabmorph} \\
\hline \hline \\[-2ex]
   \multicolumn{1}{c}{\textbf{Galaxy}} &
   \multicolumn{1}{c}{\textbf{R$_{\rm P}$  \tablenotemark{1}}} &
   \multicolumn{1}{c}{\textbf{$C$ \tablenotemark{2}}} &
   \multicolumn{1}{c}{\textbf{Class \tablenotemark{3}}} &
   \multicolumn{1}{c}{\textbf{b/a \tablenotemark{4}}} &
   \multicolumn{1}{c}{\textbf{Broad \tablenotemark{5}}} \\
   
   \multicolumn{1}{c}{} &
   \multicolumn{1}{c}{$\arcsec$} &
   \multicolumn{1}{c}{$R_{\rm{90}}/R_{\rm{50}}$} &
      \multicolumn{1}{c}{} &
         \multicolumn{1}{c}{} &
      \multicolumn{1}{c}{}
  \\[0.5ex] \hline
   \\[-1.8ex]
\endfirsthead

\multicolumn{3}{c}{{\tablename} \thetable{} -- Continued} \\[0.5ex]
   \multicolumn{1}{c}{\textbf{Galaxy}} &
   \multicolumn{1}{c}{\textbf{R$_{\rm P}$  \tablenotemark{1}}} &
   \multicolumn{1}{c}{\textbf{$C$ \tablenotemark{2}}} &
   \multicolumn{1}{c}{\textbf{Class \tablenotemark{3}}} &
   \multicolumn{1}{c}{\textbf{b/a \tablenotemark{4}}} &
   \multicolumn{1}{c}{\textbf{Broad \tablenotemark{5}}} \\
   
   \multicolumn{1}{c}{} &
   \multicolumn{1}{c}{$\arcsec$} &
   \multicolumn{1}{c}{$R_{\rm{90}}/R_{\rm{50}}$} &
      \multicolumn{1}{c}{} &
         \multicolumn{1}{c}{} &
      \multicolumn{1}{c}{}

\endhead

\multicolumn{3}{l}{{Continued on Next Page\ldots}} \\
\endfoot
%This is the footer for the last page of the table...
  \\[-1.8ex] \hline \hline
\endlastfoot

2MASX J00253292+6821442 & $\cdot \cdot \cdot$ \tablenotemark{6}& $\cdot \cdot \cdot$ & M & 0.43 & \\
2MASX J01064523+0638015 & 12.8 & 3.10 & S & 0.80 & \\
2MASX J03534246+3714077 & 8.66 & 3.03 & I & 0.56 & \\
2MASX J04440903+2813003 & 30.8 & 2.92 & I & 0.91 & \\
2MASX J05054575-2351139 & 7.14 & 3.37 & I & 0.87 & \\
2MASX J06411806+3249313 & 9.11 & 2.93 & M & 0.82 & \\
2MASX J07595347+2323241 & 15.7 & 2.80 & S & 0.49 & \\
2MASX J09112999+4528060 & 9.22 & 2.90 & I & 0.70 & \\
\end{longtable}
\end{center}

\end{document}